\documentclass[aip,apl,superscriptaddress,amsmath,amssymb]{revtex4-1}

\usepackage{graphicx}
\usepackage{algorithm}
\usepackage{algpseudocode}
\usepackage{esvect}
\usepackage{fourier}

\begin{document}

\preprint{ }

\title{ 3-Dimensional time: the physics behind quantum mechanics and unified interactions}  
\author{\normalsize Xiaodong Chen}
\email{xiaodong.chen@gmail.com} 

\date{\today}

\begin{abstract}  

If time has three dimensions, how does a particle move? This paper demonstrates that quantum physics naturally emerges from a framework
 of three-dimensional time. We present the equations governing the motion of 0-spin, 1-spin, and 1/2-spin particles within this three-dimensional time model. 
Phenomena such as quantum non-locality, spin, gauge transformation invariance, Bose-Einstein condensation, the exclusion principle, 
and Schrödinger's cat problem are shown to arise due to the presence of two additional time dimensions. We will explore how 
causality is maintained in this multi-dimensional time framework. Additionally, we will demonstrate that gravitational and electromagnetic 
fields can be unified within the 3+3 space-time Kaluza-Klein (KK) equations. Furthermore, a strong-interaction equation based on the 
$\sigma-\omega$ model is derived from the geometry of three-dimensional time for fermions. Finally, we show that gravitational, 
electromagnetic, and strong interactions can be unified within the same six-dimensional Kaluza-Klein equation.

\end{abstract}

\maketitle

\section{Introduction} \label{INTRO}
 
Over the past 120 years of developing quantum physics, several fundamental questions remain unanswered. From the double-slit interference experiment 
to Bell's inequality tests, non-locality is evident in many quantum phenomena. A central question is how to explain a particle's non-local behavior. 
Other important questions include: What exactly is spin? Why can an infinite number of particles occupy the same spatial location simultaneously 
in Bose-Einstein condensation? What is the fundamental cause of gauge transformation invariance? Do we truly understand the reason behind the infinity 
in quantum field perturbation calculations? These unresolved questions hinder our ability to explore quantum world more deeply and impede our 
understanding of the unification of field interactions. In this paper, I propose that by assuming time has three dimensions, we can directly derive the 
equations of quantum physics and quantum field theories from particle motion in 3-dimensional time. The aforementioned questions can be addressed using 
a model of 3-dimensional time combined with 3-dimensional space. In Section \ref{INTRO}, we will develop the 3-dimensional time model. Section II will describe the 
motion of single free particle in 3-dimensional time, providing the field equations for 0-spin particles. We will establish the connection between time 
vectors and wave functions and explain the double-slit interference experiment using 3-dimensional time. We will also discuss Bose-Einstein condensation, 
the uncertainty principle, and Schrödinger's cat. Session II will clarify the nature of spin. We will derive the equation of motion for 1-spin and 1/2-spin
particles.  Section III will focus on interactions, examining the discrepancy between Einstein's theory of gravity and quantum field interactions. 
We will demonstrate how 3-dimensional time physics could be the resolution of this issue. The electromagnetic interaction equation and strong interaction 
equation will be derived, and we will present a unification of the gravitational, electromagnetic fields and strong interactions using a 3+3-dimensional 
Kaluza–Klein model. We will also provide an explanation for gauge transformation invariance and describe how causality in a 1-dimensional time world 
functions within 3-dimensional time physics.

\subsection{ Nonlocality and statistical measurement result} \label{Nonlocality}

There are two fundamental  phenomena distinguish quantum physics from the classical physics: 1) Wave-packet duality. 2) Statistical result of quantum 
measurement. The double-slit experiment of electron is one of the basic experiments to show the wave-particle duality of a single particle.  
In the experiment, a electron needs to pass both slits in order to generate the interference pattern. If we try to describe the path of the electron, 
we have to admit that the electron shows up at two slits at the same time. In general, 
if we use classical physics to describe the path of wave-particle duality, we have to admit that a particle can show up many places at the same time. 
Let's carefully review the common physics principle: "A particle cannot show up two places at the same time." Assume we pick two random points on the 
path of a particle: $(x=X1, t=T1)$, $(x=X2, t=T2)$,
and the velocity of the particle is $v=(X2 - X1)/ (T2 - T1)$. When $X_2$ <> $X_1$ and $T1=T2$, 
then v will be infinite. The result is based on 1-dimensional time. If time has more than 
1 dimension and the time in the statement only refers to the clock time, then the above statement will not hold. For example, assume that time has 2-dimensions, 
let's use $t_0$ to denote the "regular" clock time dimension, $t_1$ to denote the 
second time dimension. The space-time points under two-dimensional time coordinate will be : $(X1, t_{01}, t_{11})$, $(X2, t_{02}, t_{12})$. The first index 
under t denotes the time dimension.  We have two velocities:
\begin{align} 
v_0 &= (X2 - X1)/(t_{02} - t_{01})  \nonumber \\
v_1 &= (X2 - X_1)/(t_{12} - t_{11})
\label{multipleV}
\end{align} 
When $X_2$ <> $X_1$, $t_{01} = t_{02} = t_{00}$, and $t_{12} > t_{11}$, $v_0$ is infinite, but $v_1$ is a finite number. It means that the particle 
can move from $X_1$ to $X_2$ through the second time dimension $t_{11}$ and $t_{12}$ with finite velocity $v_1$. Since we use clock to measure the first time dimension $t_0$,  
we will call $t_0$ as "clock time" through this paper. The velocity equation (\ref{multipleV}) tell us that the particle can be at two places $X_1$, $X_2$ at 
the same clock time $t_{00}$ but different coordinate value at the second time dimension.  

Now let's investigate the measurement in multiple dimensional time. We still let $t_{01} = t_{02} = t_{00}$. Assume the continuous path between $(X_1, t_{00}, t_{11})$, 
$(X_2, t_{00}, t_{12})$. Let's look at another 2-d time interval: $(t_{00}, t_{13})$ and $(t_{00}, t_{14})$. Assume the path of the motion between $(t_{00}, t_{13})$ and 
$(t_{00}, t_{14})$ is $(X_3, t_{00}, t_{13}), (X_4, t_{00}, t_{14})$. Now at clock time $t_{00}$, the particle can be at 
space coordinate interval $[X_1, X_2]$ or $[X_3, X_4]$. We have possibility to find the particle at any points in the interval $[X_1, X_2]$ and $[X_3, X_4]$, i.e. the location of the 
particle is statistical, not determinative at time $t_{00}$ because we don't have a way to measure the second time dimension. Now let's see how to calculate the probability of the 
particle's space location at clock time $t_{00}$. Since we only observe one dimensional time in macroscopic scale, if the time has multiple dimensions, the other dimensions must have small scale. 
Let's assume that the maximum length of $t_1$ is T. At clock time $t_{00}$, space coordinate of the particle is at intervals $[x_1, x_2], [x_2, x_3] ...[x_i, x_{i+1}]...[x_{n-1}, x_n]$, 
with 2nd dimensional coordinate interval $[t_{11},t_{12}]$, $[t_{12}, t_{13}]...[t_{1i}, t_{1i+1}]..[t_{1n-1}, t_{1n}]$ where $[t_{11}, t_{1n}] = [0, T]$. the probability of finding the particle 
in the sum of all intervals is 1.  the probability of finding the particle between a specific interval $[X_i, X_{i+1}]$ is |$t_{1i+1} - t_{1i}|/T$. In general, for any physical variable A 
with discrete values $A_1, A_2..A_i, A_n$. If at clock time $t_{00}$, the particle has different value $A_i$ at different 2nd dimensional time interval: $[t_{1i}, t_{1i+1}]$, then the probability 
to observe  $A_i$ is $|t_{1i+1} - t_{1i}|/T$.  
The probability is proportional to the length of the line segment of the 2nd time dimension $|t_1|$. This is a linear relationship. We know that in quantum physics, the probability is proportional to the 
the square of the magnitude of wave function $|\phi|^2$, it is not linear. Therefore, 1 extra dimension time $t_1$ is not enough to describe the statistical attribute of quantum physics. 
To derive the square of magnitude relationship, we need the 3rd time dimension $t_2$. In 3-d time $(t_0, t_1, t_2)$, the path of the particle at the $X_1$, $X_2$ and clock time $t_{00}$ becomes:
$(X_1,t_{00}, t_{11}, t_{21})$,  $(X_2, t_{00}, t_{12}, t_{22})$.  We assume the continuous path between $(X_1, t_{00}, t_{11},t_{21})$, $(X_2, t_{00}, t_{12}, t_{22})$. Let's extending the early 
discussion to 3-dimensional (3-d) time. We assume that the maximum length of $t_1$ is $T_1$, and the maximum length of $t_2$ is $T_2$. Then the probability of observe the particle at space interval $[X_1, X_2]$ 
is $|t_{12} - t_{11}|*|t_{22} - t_{21}|/T_1*T_2$. Let area segment 
\begin{equation}
S_i = |t_{1i+1} - t_{1i}|*|t_{2i+1} - t_{2i}| 
\label{timearea}
\end{equation}
Let the total area of $t_1-t_2$ is $S = T_1*T_2$. For any physical attribute A with discrete values $A_1, A_2...A_i, A_n$, the probability to observe $A_i$ is $S_i/S$.  The probability
is proportional to the area of the segment on $t_1 - t_2$ surface. We can build relationship between $S_i$ to the magnitude of wave function $|\phi|$:
\begin{equation}
\frac{S_i}{T_1T_2} = |\phi_i|^2
\label{timeAreaAndPhi}
\end{equation}

\subsection{The model of 3-dimensional time} \label{Model}

Let's give the assumptions of three-dimensional time physics:

1) Time has three dimensions. The first time dimension $t_0$ is the clock time, which can be measured by clock. the range of $t_0 $is infinite. The other two time dimension $t_1$ and
 $t_2$ are finite.
 
2) Time flow has direction. In any region of 3-d time, we can find 3 directed time flow to cover the whole region. Each directed time flow drives a motion, thus, every object 
has three independent motions. Each directed line in time flow is the projection of a worldline in 3-d time coordinate. Starting from any time point in the 3-d time coordinate, 
the particle can travel through every time point in 3-d time coordinate following the time flows if the particle has a valid physical state at that time point.     

3) We use a complex plane to represent the $t_1-t_2$ plane. Let vector $\vv{T} = t_1 + i t_2$ be a tangent vector of a timeline on $t_1-t_2$ surface,
we can find a wave-function $\phi = Re^{i\theta}$, such that the direction angle of T equals to $\theta$, and the magnitude of T: $|T| = \alpha * R$. where $\alpha$ is any real number 
and R is the amplitude of the wave-function $\phi$.    

\begin{figure}
  \includegraphics[scale=1.0]{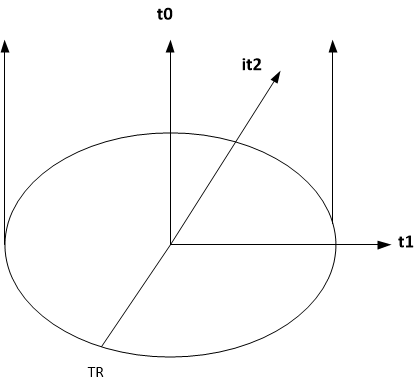}
  \caption{The three-dimensional time coordinate has cylindrical symmetry. Clock time $t_0$ is perpendicular to $t_1-t_2$ plane, $t_1-t_2$ is a complex plane}
  \label{fig:3dTime}
\end{figure}

In figure \ref{fig:3dTime}, first we draw parallel infinite lines, each line represents a $t_0$ line perpendicular to $(t_1, t_2)$ surface; then we draw a finite surface perpendicular 
to $t_0$. For symmetrical reasons, the simple 2-d finite surface is a circle, while the finite radius of the circle is $T_R$. The finite requirement of $T_R$ will bring some interesting 
physical properties at the boundary when the radius $t_r$ -> $T_R$. If vacuum has a flat geometry, it doesn't make sense that $t_r$ suddenly stops at $T_R$. To implement finite 
range of $t_1$, $t_2$, the $t_1-t_2$ plane must have a special geometry. There are many possible metrics can give limited size. Here are 3 sample metrics given finite size of time 
surface of the vacuum:
\begin{equation}
ds^2 = - dt_0^2 - dt_r^2 - (1 - \frac{t_r^2}{T_R^2}) t_r^2 dt_\theta ^2    
\label{metric_1}
\end{equation}
\begin{equation}
ds^2 = - dt_0^2 - dt_r^2 - (e^{-at_r})^2 dt_\theta ^2  
\label{metric_2}
\end{equation}
\begin{equation}
ds^2 = - dt_0^2 - dt_r^2 - T_R^2sin^2(\pi\frac{t_r}{t_R})dt_\theta ^2 
\label{metric_3}  
\end{equation}

\begin{figure}
  \includegraphics[scale=1.0]{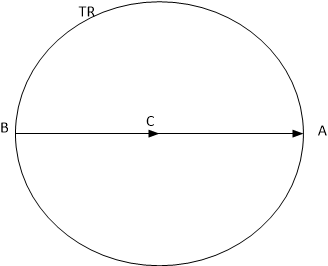}
  \caption{The arc length between A and B: $\widearc{AB} = \sqrt{(1 - t_r^2/T_R^2)} t_r \pi$. At $t_r = T_R$, $\widearc{AB}$ is 0 due to the metric (\ref{metric_1}), thus From $B \rightarrow C \rightarrow A$ forms a time loop}
  \label{fig:timeloop1}
\end{figure}

The negative sign in (\ref{metric_1})-(\ref{metric_3}) due to the time metric. When $t_r$ << $T_R$, equation (\ref{metric_1}) gives the same metric as the cylinder coordinate metric in flat geometry. 
When $t_r$ reaches the boundary $T_R$, the length of arc (circumference) is close to 0 due to  $arclength = \sqrt{(1 - t_r^2/T_R^2)} t_r d\theta $, the area segment $\Delta t_r * arclength$ also close to zero. 
Since the probability 
to find a physical attribute at any fixed clock time $t_0$ proportional to the area segment of time on $t_1 - t_2$ surface, the probability to observe any physical attribute around the boundary is also 
close to zero. Therefore, at most case, equation (\ref{metric_1}) gives an almost flat 
geometry.  The same argument can apply to geometry (\ref{metric_2}), except that $a*\hbar$ gives the mass like attribute of vacuum, we will discuss it later.  Equation (\ref{metric_3}) gives a metric similar 
to the spherical surface, it is going to be used later for 1/2-spin particle. We will use each of the three metrics throughout this paper. Let's start use (\ref{metric_1}) as the metric 
of vacuum. Fig \ref{fig:timeloop1} shows a special attribute of metric (\ref{metric_1}). At boundary $t_r = T_R$, the arc length between A and B is 0, so we can treat A and B as the same point. 
When a time flow from center C arrive A, 
it can immediately go to B then from B to C. Therefore, C->A->B->C forms a time loop. For any two events $E_1$ and $E_2$ happened on a time loop, we cannot tell which event happened first. Therefore there is no 
causality on a time loop. We know causality is important in physics, how do we keep causality when we have time loops in 3-d time? We will 
find out causality is still valid in 3-d time when we discuss interactions.

The assumption 2) tells us that time flow can be described by three independent 3-d vectors, each time vector brings an independent motion to a free particle. We use timeline to describe 
each directed line in a time flow. The last sentence in 2) defined the time continuity.  Assumption 3) builds a connection between 3-d time physics to quantum physics.  

Why do we not see the 3-d time in macroscopic scale?  Time is directly associated with motions. If all objects in the universe are static, then we don't know that time exists.
For the same reason, if the motion of objects on the 2nd and 3rd time dimensions are negligible in macroscopic level, then we won't observe the 3-d time in macroscopic world. 
The following question would be why the motion on the 2nd and 3rd time dimensions are close to zero in macroscopic scale? 
There are multiple factors: a) A macroscopic object is composed by a large number of microscopic particles. The motion of the object is a collective motion of all those particles. 
However, the different particles have different equations of motions on 2nd and 3rd time dimensions. The total motion of those particles on 2nd and 3rd time dimensions 
are cancelling each other due to different behaviors, interactions and wave-packet collapses. b) Mass is playing an important part in 3-d time geometry. 
The larger the mass of the object, the less motion on $t1-t2$ plane. c) The size of the object could also be a factor. We will discuss these factors in later chapters.   

To describe the particle's motion in 3-d time, the velocity in 3-d time would be a 3X3 matrix.
\begin{equation}
x^i = v^i_j t^j 
\label{threeDVelocity}
\end{equation}
where i=1,2,3; j=0,1,2. Repeat index indicates the sum.
If $v^i_j -> 0$ when j > 0, then the $v^i_j$ becomes the velocity vector in macroscopic world.

\subsection{The particle's trajectory in three-dimensional time} \label{Trajectory}

In classical physics, a particle's trajectory is a worldline $x(t_0)$. If time has two dimensions, a particle's trajectory will be $x(t_0, t_1)$. First, we draw 
a worldline $x(t_0)$, then at any point $x_1(t_{01})$ on worldline , we can draw a new worldline parameterized by $t_1$. As the result,  the particle's trajectory becomes a world 
surface $x(t_0, t_1)$. In 3-d time physics, at any point $x_1(t_{01}, t_{11})$, there is a worldline parameterized by $t_2$. thus, the 
world surface turns to the world field $x(t_0,t_1,t_2)$. The equation of the particle's trajectory in classical physics becomes field equations in 3-d time physics. In fact, there is  
 clear evidence of the 3-d time motion of a single particle: the motion of photon. A photon has 3 motions: the motion with $t_0$ is the classical worldline of a photon 
in relativity, the motion with $t_1$ (motion from 2nd time dimension) is a motion perpendicular to the classical worldline, described by electric field $\vv{E}$; 
the motion with $t_2$ (motion from 3rd time dimension) is a motion perpendicular to the other two motions, described by magnetic field $\vv{B}$. 
We will describe the motions of photon in detail in later chapter. 
To build equations of motion for a particle in 3-d time, we will go through three steps:
1) Find two perpendicular time flows on $t_1-t_2$ plane, the two time flows need to cover the whole plane.
2) For each timeline in the time flow, find it's mapping to a line on the space. 
3) Show that the result is a quantum wave-function.  
We are going to implement the three steps in the next two chapters on the 0-spin, 1-spin and 1/2-spin particles.

\section{How does a 0-spin particle move in three-dimensional time} \label{zeroSpin}

We use letter t as clock time, $t \equiv t_0$, and dedicate the word "speed" and letter v as classical speed v = dx/dt. 
In classical physics, the clock time t is an independent variable, space coordinate on the particle's path X= f(t) is the 
function of t. The equation of motion of an object describes the mapping 
from t to X. In 3-d time physics, since we don't have a way to accurately measure the extra time dimensions, the observation of the particle's location at clock time t 
is statistical. Therefore, we are more interested in the calculation of the probability of finding the particle at space location X. Since the area at $t_1-t_2$ plane proportional 
to the probability of finding the particle, we need to find $t_1(X)$ and $t_2(X)$ as the function of X. i.e. We need to find the inverse mapping of 3-d time to 3-d space. The equation 
(\ref{threeDVelocity}) will be rewrite as:
\begin{equation}
t^j = u^j_i x^i  
\label{threeDVelocity3}
\end{equation}
where i = 1,2,3; j=0,1,2. 
Since not all functions have inverse function, we would expect some discontinuities. 

\subsection{Equations of three proper time vectors} \label{threeProperTime}

In relativity, the proper time of a particle is the time in the particle's rest reference frame(RRF), it is a scalar. In 3-d time physics, a particle has 3 independent motions. 
Each motion has a proper time, each proper time is a vector. We denote the 3 proper time vectors as : $\vv{\tau}$, $\vv{\eta}$, $\vv{\rho}$, where $\vv{\tau}$ is the 
time vector along clock time,  $\vv{\eta}$, $\vv{\rho}$ are the time vectors along the other two time dimensions. We call worldline $\tau$, worldline $\eta$, worldline 
$\rho$ based on which proper time vector driven worldline. 

\begin{figure}
  \includegraphics[scale=1.0]{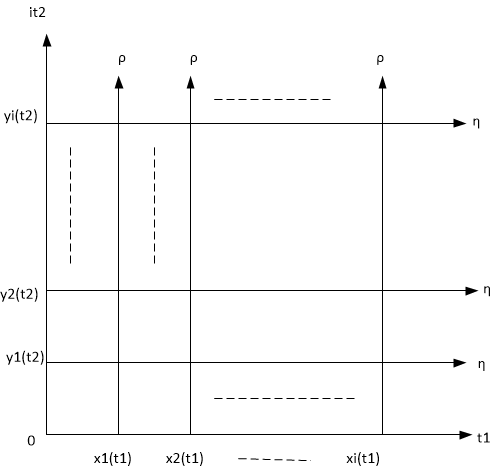}
  \caption{The paths of the particle with two additional time dimension}
  \label{fig:particlepath}
\end{figure} 

In classical physics, a Rest Reference Frame (RRF) is a reference frame (RF) moving along the object's worldline. In 3-d time, a particle has 3 independent motions, we cannot find 
a RF moving along with the particle with all 3 independent motions at the same time.  We define the RRF in 3-d time as the reference frame moving along with the particle
with the motion driven by timeline $\vv{\tau}$, i.e. in RRF, $v^{i}_0 = 0$, where $v^{i}_0$  is described by equation (\ref{threeDVelocity}). Let $\vv{q_0}$, $\vv{q_1}$, $\vv{q_2}$ 
be three unit vectors on $t_0, t_1, t_3$ axis respectively.
In RRF,  we write the equation of each proper timeline vector as:  
\begin{equation}
\vv{\tau}  = \tau \vv{q_0} + f(t_r)(cos(\omega \tau) \vv{q_1} - isin(\omega \tau) \vv{q_2}) \\ 
\label{tauMotion}
\end{equation}
\begin{equation}
\vv{\eta} = \eta cos(\theta_0) \vv{q_1} + i\eta sin(\theta_0) \vv{q_2} \\ 
\label{etaMotion}
\end{equation}
\begin{equation}
\vv{\rho}  =  - \rho sin(\theta_0) \vv{q_1} + i\rho cos(\theta_0) \vv{q_2}   
\label{rhoMotion}
\end{equation}
where $\omega$ is the rotation frequency, $\theta_0$ is initial phase angle, $f(t_r)$ is the distance to the center of the rotation, which will be the function of $\eta$ and $\rho$. 
Since we treat $t_1-t_2$ plane as a complex plane, we can rewrite the 
above equations as:
\begin{equation}
\vv{\tau}  = \tau \vv{q_0} + f(t_r)e^{-i\omega \tau} \vv{q_\perp}  \\ 
\label{tauMotion2}
\end{equation}
\begin{equation}
\vv{\eta}  = \eta e^{i\theta_0} \vv{q_\perp} \\ 
\label{etaMotion2}
\end{equation}
\begin{equation}
\vv{\rho} = i \rho e^{i\theta_0} \vv{q_\perp}  
\label{rhoMotion2}
\end{equation}

where $\vv{q_\perp}$ represents the unit vector on $t_1 - t_2 $ plane which perpendicular to $t_0$ axis. 
At a fixed clock time $t_0 = 0$, we choose $\theta_0 = 0$. then $\vv{\eta} =  \eta \vv{q_\perp}$, $\vv{\rho} =  \rho \vv{q_\perp}$
Since in classical physics, a free particle's velocity is constant. It is reasonable to assume that in 3-d time, the velocity of 0-spin free particle in RRF is also constant.  
Let's choose space coordinate direction such that 
\begin{eqnarray}
\frac{dx}{d\eta} = v_1 \nonumber	\\
\frac{dy}{d\eta} = \frac{dz}{d\eta} = 0 \nonumber	\\
\frac{dy}{d\rho} = v_2	\nonumber	\\			
\frac{dx}{d\rho} = \frac{dz}{d\rho} = 0			
\label{etaVx}
\end{eqnarray}
 where $v_1$ and $v_2$ are constant velocities.

 Fig \ref{fig:particlepath} shows the time flows in RRF at clock time $t_0 = 0$. Each line of time flow of $\eta$
 parallel to $t_1$ axis, the different $\eta$ line starts from different $t_2$, then move along $t_1$ direction; each line of time flow of $\rho$ parallel to $t_2$ axis, the 
different $\rho$ line starts from different $t_1$, then move along $t_2$ direction. Along timeline $\eta$, the path of the particle parallel to x axis; 
 along timeline $\rho$, the path of the particle parallel to y axis. Space coordinates corresponding to each time coordinates are marked on the $t_1,t_2$ axis.
 At time $(0,t_1, 0)$, the particle is at space point $(x1,0)$, the particle can move from $(x1,0)$ 
 to (x1,y1) through  timeline $\rho$. In general, there are many paths from (x1,y1) to reach (xi,yi).  For example, the particle can move from (x1,y1) to (x1,yi) 
 along $\rho$ timeline, then move from (x1,yi) to (xi,yi) along $\eta$ timeline. Or the particle can move from (x1,y1) to (xi,y1) along $\eta$ timeline first, 
 then move from (xi,y1) to (xi,yi) along $\rho$ timeline. Or go through any directed combination of $(x_i, y_j)$.. the probability of finding the particle in grid 
 $\Delta x$ $\Delta y$ is proportional to $\Delta t_1 \times \Delta t_2$. space paths of $\eta$ and $\rho$ in space coordinate are not necessarily perpendicular to each other. 
 They can form any angle. In fact, in one dimensional space case,  both $\eta$ and  $\rho$ can map to the same x direction. 
 
 Both $t_1$ and $t_2$ are finite, so does each timeline on $t_1-t_2$ plane. Let the range of $\eta$ be $[T_A, T_B]$, 
then using (\ref{etaVx}), the range of x is $[x_A, x_B] = [v_1 T_A, v_1 T_B]$. From Fig \ref{fig:timeloop1}, the arc length between A 
and B is 0. Therefore, when the particle arrives $x_A$, it will immediately go to $x_B$ with the same constant velocity $v_1$ and continue moves from $x_B$ to $x_A$, it forms a loop. 
We can apply the same analysis on the motion of $\rho$ on y direction. 

\begin{figure}
  \includegraphics[scale=1.0]{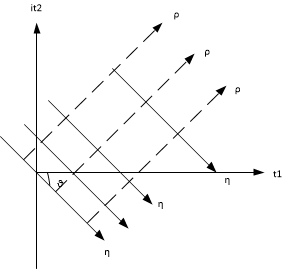}
  \caption{The direction of timeline vector $\eta$ and $\rho$ rotate $\theta$ degree clockwise when $t_0$ changes, $\theta = \frac{m_0c^2}{\hbar} t_0$ }
  \label{fig:timerotation}
\end{figure}

Equation (\ref{tauMotion}) is a helix equation. It shows that in RRF, when $t_0$ changes, the whole $t_1-t_2$ surface of the particle rotates with frequency $\omega = m_0 c^2 /\hbar$, where $m_0$ is the 
particle's rest mass and c is the speed of light. Fig \ref{fig:timerotation} shows that when $t_0 = \tau > 0$,  the direction of $\eta$ at $t_1-t_2$ plane changes with the angle 
$\theta = \frac{m_0c^2}{\hbar} \tau$. Equation (\ref{tauMotion}) is a very important equation because it is the oscillation of the wave-function. It shows that in RRF, as long as the particle has energy,
the whole 3-d time coordinates will rotate with $\tau$. i.e. The direction of all the time vectors on $t_1-t_2$ plane will change with $\tau$.

\subsection{Wave speed and group speed } \label{waveSpeedGroupSpeed}

How to connect 3-d time motions to mass-wave function? We know that a wave is formed by two parts: a) at each location, the particle oscillates with time. b) the oscillation 
propagates through space with wave-speed w.  From equation (\ref{tauMotion2}) , we see that at any point, time vectors oscillates with $\tau$,  therefore, a) is covered. 
Now we need to find out how does motions driven by $\eta$ and $\rho$ propagate the oscillations. 

In quantum physics, wave speed
\begin{equation}
 w = \lambda  \nu 
 \label{waveSpeed0}
 \end{equation}
 where $\lambda$ is wave-length, $\nu$ is frequency. Substitute de Broglie equations   
\begin{eqnarray}
\lambda = \frac{h}{mv}  \\
\nu = \frac{E}{h} = \frac{mc^2}{h}
 \label{debroglie}
 \end{eqnarray}
into (\ref{waveSpeed0}), we have 
\begin{equation}
w = \frac{c^2}{v}
 \label{waveSpeed2}
 \end{equation}
where c is speed of light, v is speed of the particle. Since v << c, we have wave speed w >> c. In traditional explanation, v is the group speed and w is phase speed. 
Since in relativity, it is not possible to have the wave propagates with speed larger than speed of light, how do we understand that the particle's phase speed >> c?  We  
will show that the phase speed w >> c is a direct result of 3-d time physics.

In the previous section, we see that in RRF, the particle moves on x axis with timeline $\vv{\eta}$ as shown in Fig 3. This motion happens without changing the clock time.
For any two points $x_1, x_2$ on $\vv{\eta}$ path, $\Delta t_0 = 0$. Using the definition of speed in classical physics: $w = dx / dt_0$, the particle's speed 
$w = |x2 -x1|/\Delta t_0 = \infty$. i.e. the particle propagates the rotation along x-axis with infinite speed. 

An observer moves with speed v relative to the particle's RRF, then the observer sees that the particle moves with speed v along worldline $\tau$. Using Lorentz transformation, 
the clock time at  $x_1,x_2$ are also changed. From the observer's view, the particle arrives $x_1, x_2$ on worldline $\eta$ at different clock time. It means worldline $\eta$ 
obtain a speed on $t_0$ direction due to Lorentz transformation. We have 
\begin{align}
\Delta t &= \gamma (\Delta t' + v \frac{\Delta x'}{c^2})  \nonumber \\
\Delta x &= \gamma (\Delta x'  + v \Delta t')
\label{etaLorenzT}
\end{align}
where $\Delta t'$ is the clock time difference along the motion of $\eta$ in RRF. Therefore, $\Delta t' = 0$.  Substitute $\Delta t' = 0$ into 
(\ref{etaLorenzT}), and then divided the second equation by the first equation,  the speed of worldline $\vv{\eta}$ on $t_0$ direction in observer's RF becomes 
\begin{equation}
w = \frac{\Delta x} {\Delta t} = \frac{c^2}{v}
\label{etaWaveSpeed}
\end{equation}
Equation (\ref{etaWaveSpeed}) equals equation (\ref{waveSpeed2}). thus, we show that the phase speed of the particle is from the motion of $\vv{\eta}$. This result gives another 
prove that the motion of 2nd time dimension exists. The same can apply to the motion of $\rho$. 
The motion from 2nd time dimension and 3rd time dimension propagates the matter-wave in space, the motion of the 1st time dimension provides the oscillation.  

Let's write timeline vectors  $\vv{\eta}$,  $\vv{\rho}$ as function of $\tau$:
\begin{equation}
\eta (\tau) = \eta e^{-i\omega \tau} 
\label{tau1}
\end{equation}
\begin{equation}
\rho (\tau) = i\rho e^{-i\omega \tau} 
\label{rho1}
\end{equation}
i in (\ref{rho1}) is due to $\rho$ starting with $t_2$ axis. 
Equation (\ref{tau1}) and $\tau \vv{q_0}$ forms a helicoid. Compared to the helix equation:
\begin{eqnarray}
x = r cos (\alpha) \; \; 
y = r sin (\alpha) \; \;
z = s \alpha       
\label{Helicoid}
\end{eqnarray}
We have 
\begin{equation}
r = \eta, \; \; \alpha = -\omega \tau , \; \;  s = \frac{1}{\omega}
\label{Helicoid2}
\end{equation}
$\rho$ and $\tau \vv{q_0}$ also forms a helicoid.

\subsection{ Wave function, Bose-Einstein condensation and uncertainty principle } \label{WaveFunctionAndBECondensation}

The definition of wave-function depends on the choice of representation. How do we define wave-function $\psi(x)$ in space coordinate representation in 3-d time terminology? 
In Fig \ref{fig:particlepath}, the probablity to find the particle in $\Delta x \Delta y$ is proportional to $\Delta t_1 \times \Delta t_2$. Since $|\psi(x)|^2$ is the probability
of finding the particle at x, we need to let $\Delta x, \Delta y, \Delta t_1, \Delta t_2 \rightarrow 0$. Therefore, $\psi(x)$ should be proportional to the tangent of timeline
$\eta$ and $\rho$. 

Let's write timeline $\eta$, $\rho$ as function of x and $\tau$: $\eta \equiv \eta(x, \tau)$, $\rho \equiv \rho(x, \tau)$. Let $\vv{u_{\eta}}$, $\vv{u_{\rho}}$ be tangent vectors of 
$\eta(x, \tau)$ and $\rho(x, \tau)$, then 
\begin{equation}
\psi(X,t) = c_1 \vv{u_{\eta}}  
\label{wavFunctions}
\end{equation}
\begin{equation}
\psi(X,t)^* = c_2 \vv{u_{\rho}}*  
\label{wavFunctionsConjugate}
\end{equation}

Use equation (\ref{etaVx}), we have 
\begin{equation}
 \vv{u_{\eta}} = \frac{1}{v_1}e^{-i\omega \tau} \vv{e_x} 	 
\label{wavFunctionsDirection1}
\end{equation}
\begin{equation}
\vv{u_{\eta}} = i\frac{1}{v_2}e^{-i\omega \tau}	\vv{e_y} 
\label{wavFunctionsDirection2}
\end{equation}
where $\vv{e_x}$ is unit vector on x direction, $\vv{e_y}$ is unit vector on y direction. 

 Substitute (\ref{wavFunctionsDirection1}) and (\ref{wavFunctionsDirection2}) into (\ref{wavFunctions}) (\ref{wavFunctionsConjugate}), 
$\psi(X,t) = \frac{c_1}{v_1} e^{-i\omega \tau}$, $\psi(X,t)^* = -\frac{c_2}{v_2} e^{i\omega \tau}$. If we choose $c_1 = R v_1$, $c_2 = -R v_2$,
where R is normalization constant, then we get wave-function of a plane wave in quantum physics in RRF. 

In a general RF, the particle moves with momentum p > 0. From Lorentz transformation: 
$\tau = \gamma (t - v/c^2 x) => m_0c^2\tau = Et - px$.  To simply the argument, Let's assume that $p$ moves along x space direction. Space direction of 
worldline $\eta$ can be any direction, let's say that the particle moves from $(x_1,y_1)$ to $(x_2,y_2)$ on worldline $\eta$. We can decompose space direction of worldline $\eta$
to two components: one component is parallel to $\vv{x}$: $\eta_{x}$ , the other component $\eta_{y}$ is perpendicular to $\vv{x}$: $\eta_{y}$ . At a fixed clock time t, the time vector angle
of worldline $\eta$ only changes along x direction, not on y direction. Therefore, only $\eta_{x}$ has contribution to the wave-function. Then in general RF, 
the strict definition of (\ref{wavFunctions}) (\ref{wavFunctionsConjugate}) is:
\begin{equation}
\psi(X,t) = (\vv{e_{\eta}} \cdot \vv{e_{\tau}}) c_1 |u_{\eta}| \vv{q_\eta} \vv{e_\tau}   
\label{waveFunction0}
\end{equation}
\begin{equation}
\psi(X,t)^* = (\vv{e_{\rho}} \cdot \vv{e_{\tau}}) c_2 |u_{\rho}| \vv{q_\rho}^* \vv{e_\tau} 
\label{waveFunction1}
\end{equation}
\begin{equation}
\vv{q_\eta} = e^{\frac{-i}{\hbar} (Et - px) } 	
\label{waveFunction2}
\end{equation}
\begin{equation}
\vv{q_\rho} = i\vv{q_\eta}  
\label{waveFunction3}
\end{equation}
where $\vv{e_{\eta}}$($\vv{e_\rho}$) is the unit vector pointing to space direction of the $\eta$ ($\rho$) motion, $\vv{e_{\tau}}$ is space unit vector of worldline $\tau$, which is the 
same as the direction of momentum $p$ for a longitudinal wave. The time direction of the wave function is determined by $\tau$ due to the time oscillation from $\tau$.
The space direction of the wave function is also determined by $\tau$ due to Lorentz transformation. Space direction of $\eta$ ($\rho$) is decomposed to two components,
the component parallel to $\vv{e_{\tau}}$ propagates the wave. Therefore, it is included in the wave-function; the component perpendicular to $\vv{e_{\tau}}$ contribute to the width of the wave. 

One issue of this definition is that 
if $\vv{e_{\eta}}$ or $\vv{e_{\rho}}$ perpendicular to $\vv{e_{\tau}}$, the inner product in (\ref{waveFunction0}) (\ref{waveFunction1}) would be zero, then we don't have $\psi(X,t)$ 
or $\psi(X,t)^*$. Since we cannot observe the space direction of $\eta$ and $\rho$ motion, there is very little possibility that $\vv{e_{\eta}}$ or $\vv{e_{\rho}}$ exactly perpendicular 
to $\vv{e_{\tau}}$. As long as there is a small spatial component of the directions $\eta$ and $\rho$ along the direction of $\vv{e_{\tau}}$, their inner product will not be zero. 
In practice, the inner product of the spatial directions with the constants $c_1$ and $c_2$ will be absorbed into the normalization constant of the wave function, so the 
equations (\ref{waveFunction0}) and (\ref{waveFunction1}) remain valid. In real experiments involving a large number of particles, even if a few particles 
have $\vv{e_{\eta}}$ or $\vv{e_{\rho}}$ perpendicular to $\vv{e_{\tau}}$, their effect can be neglected. Additionally, the probability of finding a particle is the product of 
$P_{\eta}$ and $P_{\rho}$, where $P_{\eta}$ and $P_{\rho}$ are the probabilities of finding the particle on timelines $\eta$ and $\rho$, respectively. 
The two probabilities are independent, so the ratio between $\vv{e_{\eta}}$ and $\vv{e_{\rho}}$ along the $\tau$ direction does not affect the probability. However, in special cases, 
the difference between the motions along $\eta$ and $\rho$ might significantly impact the quantum effects. If such a case arises, we would need to investigate whether current quantum 
physics or three-dimensional time physics provides the correct description.
 
We only discussed the motions of a single particle in 3-d time physics so far. Let's look at the motion of many particles. Considering two particles a1 and a2, both satisfy the 
equations (\ref{etaVx}) with different initial condition along $\eta$. For particle a1, the initial location x is $x = 0$ at 3-d time (0,0,0); for particle a2, 
the initial location x is $ x = -a $  at time (0,0,0). Then a1 will arrive x = a at $(0, \frac{a}{v1},0)$, a2 will arrive x = a at $(0, \frac{2a}{v1},0)$, where v1 is velocity of $\eta$
from equation (\ref{etaVx}). The particles a1 and a2 will never run into
each other. But from the view of 1-dimensional time, the two particles a1 and a2 occupy the location x=a at the same clock time t=0.  We can apply the same argument on 
infinite number of particles. From the view of 1-dimensional time physics, we are seeing infinite number of particles occupy the same space point x=a at the same time t=0, 
but from the view of 3-d time physics, each particles arrives x=a at different $t_1$. This is the physics behind Bose-Einstein condensation. 

Equation (\ref{etaVx}) gives the 3-d motions of the particle in RRF, the momentum $p = mv = 0$. It tells us that when the particle's momentum is 0, the particle can still
move to different location with $\eta$ and $\rho$. The location of the particle is uncertain at a fixed time and momentum equals 0. The result can be extended to any fixed momentum p using Lorentz
transformation. This is the physics behind the momentum and position uncertainty principle. The same can apply to energy and time uncertainty. If a particle is involved in two events. 
Event A happened at x=a, event B happened at x=b. We don't know the change of $t_0$ between event A and B, because the particle can go from a to b through motion $\eta$, motion $\rho$ or 
motion $\tau$, or the combination of three motions. Different motion paths have different change of time $\Delta t_0$, therefore, $\Delta t_0$ is uncertain.

\subsection{ Time-vector, superposition principle and double-slit experiment } \label{DOUBLESLIT}

In Fig \ref{fig:particlepath}, each $\eta$ timelines parallel each other. What if due to some initial conditions, two timelines $\eta_1$ and $\eta_2$ are not parallel, they cross each other on $t_1-t_2$ surface?
Since each timeline is a vector, at the crossing point, the two timeline should follow the math of vector addition, it will combine to a new vector $\eta_3$:
$\eta_3 = \eta_1 + \eta_2$. In RRF, using equation (\ref{wavFunctions}), we have 
\begin{equation}
\psi_1 + \psi_2 = c_1 (\vv{u_{\eta}}_1 + \vv{u_{\eta}}_2)
\label{superposition1}
\end{equation}
We see that the superposition principle of wave-function is coming from the vector addition of time vectors.  If at a space location X,  ${u_{\eta}}_1$ 
and ${u_{\eta}}_2$ have the same magnitude but opposite directions, $\vv{u_{\eta}} _1 + \vv{u_{\eta}} _2 = 0$, then the particle cannot show up at X, 
the probability of finding the particle at X is 0. 

\begin{figure}
  \includegraphics[scale=1.0]{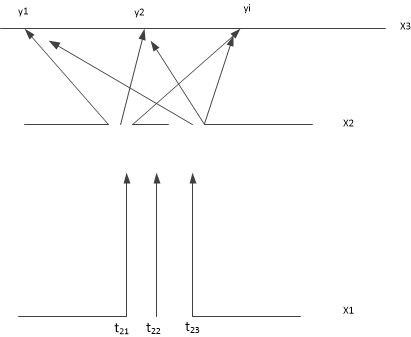}
  \caption{A single particle passes double-slit, each timeline starts with different $t_2$ value at X1. When they reache the double-slit, the different timelines scatter to various directions at X2,
	and form an interference pattern at X3}
  \label{fig:doubleslit}
\end{figure}

Fig \ref{fig:doubleslit} shows a double-slit experiment. A particle is emitted at $x=x_1$ with momentum p on x direction. Fig \ref{fig:doubleslit}  draws the motion of timeline $\eta$ at x direction. 
When $x = x_1$, all timelines start
at $t_1 = 0$ but having different initial $t_2$ = $t_{21}, t_{22}, . t_{2i}$.   At $x = x_2$, the particle's timelines pass the double-slits, half of the lines pass the left slit, the other half 
pass the right slit. The particle's $\eta$ timelines are split to two branches. At each branch, the slit scatters each worldline to different momentum directions, the momentum $\vv{p}$
on y direction is not zero anymore. When the lines meet at $x=x3$, each timeline vector has different time direction, we need to do time vector addition at $x_3$. The same argument 
apply to $\rho$ timelines. Due to the vector addition, the time vector $u_{\eta}$ and $u_{\rho}$  has maximum magnitude at some $y_i$ , and minimum magnitude at some other $y_j$. Since the probability of 
finding a particle at $(x_3, y_i)$ proportional to $|u_{\eta}(x_3,y_i) \times u_{\rho}(x_3,y_j)|$, we get the interference pattern at $x_3$.

\subsection{ Time branches, time jump, wave-packet collapse and Schrödinger's cat} \label{TIMEBRANCHS}

\begin{figure}
  \includegraphics[scale=1.0]{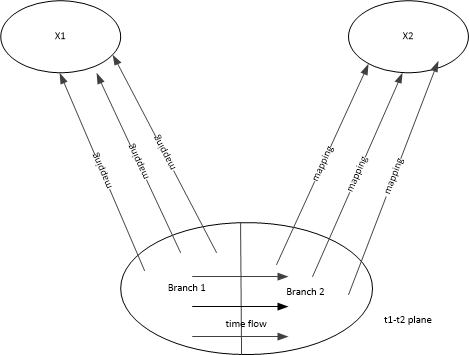}
  \caption{A particle's $t_1-t_2$ plane splits into two branches. X1, X2 are two space regions far away from each other. Branch1 maps to X1, Branch2 maps to X2, the horizental time flow 
  travels between two branches}
  \label{fig:timebranch}
\end{figure}

In the analysis of the double-slit experiment, we observe that a particle's time flow is split into two branches by the two slits. Afte the double-slit, the two branches combined into one. What if 
we add a infinite wall between two slits? Then the two branches will keep separated. In fact, at any fixed clock time t, the particle's 
time flows on $t_1-t_2$ plane can be split to many branches, and each branch has different physical properties. For example, $t_1-t_2$ can be split into n branches, each branch 
has different value of energy $E_i$. Start at any point at the $t_1-t_2$ plane, the particle can go through all values of $E_i$ through the time flows. Considering a particle has 
two spin states: left and right. Each spin state belongs to one time branch. At the beginning of clock time $t=0$, the two time branches are next to each other at a space location. 
Then the worldlines of two branches travel to opposite directions, after the clock time passed t >> 0, the opposite motions reach two completely separated  far away space locations $X_1$ and $X_2$. 
Fig \ref{fig:timebranch} shows the mapping between space location X1, X2 and time branches1 and 2. The particle's each worldline is composed of space coordinate and time coordinate. 
Although the two space location X1 and X2 are separated  at t >> 0, the projection of worldlines on the $t_1-t_2$ plane are two adjacent branches. The particle travels from branch 1 
to branch 2 through time flows. i.e. The particle can jump from X1 to X2 through time flow at t without passing any space points in between. We call it: time jump.

Why do we never observe a particle showing up at two places at once? Starting from clock time t=0 to t >> 0, the particle goes through a adiabatic process. Using the terminology of Feynman diagram,
when the particle's worldlines smoothly separated to two opposite directions, there is no interactions causing the particle annihilation and recreation during the process, 
thus, the 3-d time flows are smooth. When the measurement happened at clock time $t_0 = T_0$ >> 0 , the measurement destroyed the particle's time flow, the particle 
will go through annihilation and recreation. Suppose the annihilation happens at branch 1 and space location X1. After annihilation, the particle's time flow is interrupted. When 
the particle recreated and clock time moves to $t_0 = T_1$, the new time flow at new $T_1$ will be based on the physical properties at branch 1, the history (memory) of brach2 is gone, 
branch 1 will be expanded to branch 2 with the physical properties in branch 1. Thus, after recreating, the whole $t_1-t_2$ will all map to X1, the particle will not show up at X2 anymore. 
We call it: Wave-packet collapse. 

Now we can see that three-dimensional time physics is not multiverse. Every timeline happens at the same universe, the particle goes through 
every timeline at any clock time t, and measurement will cause the particle annihilation and recreation then cause the timelines collapsed. In macroscopic world,
every object is composed by enormous amount of particles. Interactions happens among those particles caused the wave-package collapse (timelines collapse) 
at every clock time t and every location x. That's the main reason we don't see three-dimensional time motion in macroscopic world.  

Now let's talk about Schrödinger's cat. When the cat is dead, it is annihilated, the wave-packet has already collapsed even without measurement. The cat will keep dead.
The cat can have two states: health and sick at the same time,  but cannot be both live and dead. 

One question could raise that if the particle is in different energy level at the same clock time in different $(t_1, t_2)$ time branch, why the particle doesn't jump from 
higher energy level to lower energy level in the high energy branch? To answer this question, we need to understand that lowest energy principle is based on the principle of 
least action on timeline $\tau$. In each time branch, along each $\tau$, there is only one energy, each worldline $\tau$ doesn't go through different energy levels. 
The time jump between energy branches are not through timeline $\tau$. The energy value on each time branch is still local minimum on $\tau$ motion, 
it still obeys the principle of least action, that's why no energy jump happens. 
 
\subsection{ 6-dimensional Einstein field equation and Klein-Gordon equation} \label{zeroSpinField} 

In general relativity, a star generates a curved local space-time due to its mass. In 3-d time physics, we assume that a particle generates the curved 3-d time due to its 
energy. Since "local" often refers to space location, instead of calling the particle's curved time geometry as "local" field, we call it the particle's "self" time field. 
Let's write the metric of the "self" time field for a free spinless particle:   
\begin{equation}
ds^2 = - dt_0^2 - dt_r^2 - (f(t_r) e^{\frac{-i}{\hbar}(Et_0 - px)})^2  {dt_\theta} ^2    
\label{cylindar_metric_0}
\end{equation}

Here we use the cylinder coordinate. Compared to the regular cylinder metric, the extra term $e^{\frac{-i}{\hbar}(Et_0 - px)}$ displays the fact that the particle's whole time field 
rotating with $t_0$, the negative sign due to the metric is for 3-d time instead of 3-d space. Let 
\begin{equation}
\phi = f(t_r) e^{\frac{-i}{\hbar}(Et_0 - px)}
\label{phi_0}   
\end{equation}
And includes space metric, we have 6-dimensional metric to describe the particle's self time-space geometry field: 
\begin{equation}
\left( \hat{g}_{AB} \right) = \left( \begin{array}{cc}
   g_{\alpha\beta} \; \; \; \; \; \;  \; \; \; \; \; \; \\
  \; \; \; \; \; \; \; \; -\phi^2 \; \; \; \; \; \; \; \; \\
   \; \; \; \;  \; \; \; \; \; \; \; \; \; \; -1 \\ 
   \end{array} \right) 
\label{6dMetric_0}
\end{equation}
where $g_{\alpha\beta}$ is four-dimensional metric tensor. The four-dimensional metric signature is taken to be $(- \, + \, + \, +)$, 
and the indices for 6-dimensional time-space to be 0,1,2,3,4,5, $x_0 = t_0; x_1,x_2,x_3$ are space coordinates, $x_4 = t_{\theta}, x_5 = t_r$,
we choose the speed of light c = 1.  

We extend the original Kaluza-Klein(KK) theory \cite{Kal21} \cite{Kle26a} to 3-d time + 3-d space.  
The 6-dimensional Einstein equations keep the same format as 4-dimensional \cite{Overduin}:
\begin{equation}
\hat{G}_{AB} = \kappa \hat{T}_{AB} \; \; \; ,
\label{5dEFE1}
\end{equation}
where $\kappa$ is a constant; $\hat{T}_{AB}$ is 6-dimensional energy momentum tensor, 
$\hat{G}_{AB} \equiv \hat{R}_{AB} - \hat{R} \, \hat{g}_{AB} / 2$
is the Einstein tensor, $\hat{R}_{AB}$ and
$\hat{R} = \hat{g}_{AB} \hat{R}^{AB}$ are 
the 6-dimensional Ricci tensor and scalar respectively,
and $\hat{g}_{AB}$ is the 6-dimensional metric tensor, A, B.. run over 0,1,2,3,4,5 .

The 6-dimensional Ricci tensor and Christoffel symbols are defined
in terms of the metric exactly as in four dimensions:
\begin{eqnarray}
\hat{R}_{AB}        & = & \partial_C \hat{\Gamma}^C_{AB} -
                          \partial_B \hat{\Gamma}^C_{AC} +
                          \hat{\Gamma}^C_{AB} \hat{\Gamma}^D_{CD} -
                          \hat{\Gamma}^C_{AD} \hat{\Gamma}^D_{BC} 
                          \; \; \; , \nonumber \\
\hat{\Gamma}^C_{AB} & = & \frac{1}{2} \hat{g}^{CD} \left( 
                          \partial_A \hat{g}_{DB} +
                          \partial_B \hat{g}_{DA} -
                          \partial_D \hat{g}_{AB} \right) \; \; \; 
\label{6dChristRicci}
\end{eqnarray}
where A, B.. run over 0,1,2,3,4,5. 
We concentrate on quantum field equations, so ignore gravity field, then 
$g_{\alpha\beta} = diag(-1,1,1,1)$, $g_{55} = -1$, $g_{44} = -\phi^2$,
with conditions: 
\begin{eqnarray}
\partial_{4} \phi = 0  \;, \; \;  \;
\partial_{5} \phi =  -\frac{m_0}{\hbar} \phi \; \; \;  \; 
\label{6DCondition} 
\end{eqnarray}
where $\hbar$ is Planck constant, $m_0$ is rest mass of particle, or equivalently
\begin{equation}
\phi =  e^{-\frac{m_0}{\hbar} t_r} \psi 
\label{6DCondition_1} 
\end{equation}
where $\psi$ is a wave-function. We let $f(t_r) = e^{-\frac{m_0}{\hbar} t_r}$. Compared to metric (\ref{metric_2}),  $e^{-\frac{m_0}{\hbar} t_r}$ 
provides the boundary condition of $t_1-t_2$ plane. When $t_r$ increases , the arc length of the circle shrinks exponentially.
Note the time metric for vacuum could still be (\ref{metric_1}), the particle's mass contributes the additional exponentially 
shrink. Also we have $g_{AB} g^{AB} = \delta_{AB} $, that makes $g^{44} = -\frac{1}{\phi^2} $, and $g^{\alpha \beta} = diag(-1,1,1,1)$. 

Using metric (\ref{6dMetric_0}), we get following non-zero Christoffel symbol:
\begin{eqnarray}
\hat{\Gamma}^{4}_{4 \alpha } = \frac{1}{2} g^{44} \partial_{\alpha} g_{44} \;, \; \; \; \; \; \; \;
\hat{\Gamma}^{4}_{{\alpha} 4} =  \hat{\Gamma}^{4}_{4 \alpha } \nonumber \\
\hat{\Gamma}^{\alpha}_{44}  =  -\frac{1}{2}g^{\alpha\alpha}\partial_{\alpha}g_{44}  \;,  \; \; \; 
\hat{\Gamma}^{4}_{45} =  \frac{1}{2}g^{44}\partial_{5}g_{44} \nonumber \\
\hat{\Gamma}^4_{54}  =  \hat{\Gamma}^{4}_{45}  \;,  \; \; \; \;
\hat{\Gamma}^{5}_{44}  =  -\frac{1}{2}g^{55}\partial_{5}g_{44} 
\label{5dChristoffel}
\end{eqnarray}

Substitute $g_{44} = -\phi^2$ and (\ref{5dChristoffel}) in (\ref{6dChristRicci}), Ricci tensor becomes:
\begin{eqnarray}
R_{\alpha\beta} =  -\frac{1}{\phi}\partial_{\alpha}\partial_{\beta}\phi \nonumber \\
R_{\alpha 5} = R_{5\alpha} =  -\frac{1}{\phi}\partial_{\alpha}\partial_{5}\phi \nonumber \\
R_{55} = -\frac{1}{\phi}\partial_{5}\partial_{5}\phi  \nonumber \\
R_{\alpha4} = R_{4\alpha} = 0 \nonumber \\
R_{44}  = \phi\partial^{\alpha} \partial_{\alpha}\phi-\phi\partial^{5} \partial_{5}\phi 
\label{5dRicci}
\end{eqnarray}
Let $R_{44} = 0$, which means no enegery-momentum tensor $T_{44}$. We get 
\begin{equation}
\phi(\partial^{\alpha} \partial_{\alpha}\phi +\partial^{5} \partial_{5}\phi) = 0
\label{RicciScalar} 
\end{equation}

From equation (\ref{6DCondition_1} ), $\partial^{5} \partial_{5}\phi = -\frac{m^2}{\hbar^2} \phi$. Equation (\ref{RicciScalar}) becomes Klein-Gordon
equation: 
\begin{equation}
\partial^{\alpha} \partial_{\alpha}\phi- \frac{m_0{^2}}{\hbar^2}\phi = 0
\label{KleinGordon} 
\end{equation}
We take 4-dimensional metric $(- \, + \, + \, +)$. 

$\phi$ is not a wave-function. From (\ref{6DCondition_1}), $\phi$ contains an extra term $e^{-\frac{m_0}{\hbar} t_r}$, the field equation obtained mass from this term.
Then $\frac{1}{\phi}$ removes $e^{-\frac{m_0}{\hbar} t_r}$ after the field obtain the mass, equation ($\ref{KleinGordon}$ ) then turned to Klein-Gordon equation for wave function
$\psi$. It has the advantage to use $\partial^{5} \partial_{5}\phi$ instead of $\frac{m_0{^2}}{\hbar}\phi$. $\partial_{5}$ will ensure gauge transformation invariance for the mass term,
we will discuss it in later chapter. 

The Einstein equations (\ref{5dEFE1}) become:
\begin{equation}
  R_{AB} = \kappa \hat{T}_{AB} 
\label{6dMSTensor_0}
\end{equation}
If we let
\begin{equation}
  \kappa = (\frac{1}{\hbar})^2 
\label{kappa}
\end{equation} 
Using equations (\ref{5dRicci}), we get  
\begin{equation}
 -\frac{1}{\psi}\hbar^2 \partial_{\alpha}\partial_{\beta}\psi = T_{\alpha\beta} 
\label{energyMomenum}
\end{equation} 
If $\psi$ is a plane wave function, then equation (\ref{energyMomenum}) becomes   
\begin{equation}
 p_{\alpha}p_{\beta} = T_{\alpha\beta} 
\label{energyMomenum2}
\end{equation} 
From the above equation, we see $\frac{1}{\hbar}$ 
plays the similar role as $4\pi G$ 's role in 4-dimensional Einstein equation, where G is gravitational constant

\subsection { Section summary} \label{SectionISummary}

In summary, a free 0-spin particle has three motions, each associated with timeline vector $\tau$, $\eta$, $\rho$ respectively. Each motion has constant velocity. Motion $\tau$ provides a 
rotation of the whole $t_1-t_2$ plane, which changes the direction of  $\eta$, $\rho$. This rotation doesn't associate with space motion, it provides the phase oscillation of wave-function. 
The rotation is due to the distorted 3-d time metric created by energy of the particle in (\ref{cylindar_metric_0}). Superposition principle comes from the vector addition of timelines; 
other properties of 3-d time physics is due to the nature of 3-d time.

\section{ Equation of motion of 1-spin massless particle and 1/2-spin massive particle } \label{EquationOfMotion1andhalf}

\subsection{The equations of motion for photon} \label{PHOTON}

A particle's worldlines (world field) are driven by three separate motions: $\vv{\tau}$, $\vv{\eta}$, $\vv{\rho}$. Photon is an ideal particle to displays such behavior. 
A photon has three motions: a classical motion with clock time $t_0$ at the speed of light, two motions $\vv{E}$, $\vv{B}$ perpendicular to the direction of the path of the 
classical motion. The probability to observe of a physical property of a particle is proportional to $|\Delta t_1  \times \Delta t_2|$. 
We can see that from the definition of Poynting vector of the photon:
\begin{equation}
 \vv{P} = \frac{1}{4\pi c} \vv{E} \times \vv{B} 
\label{PoyntingVector}
\end{equation}  
where $\vv{P}$ describes the momentum density of the photon. 

With all above promising aspects, how do we define the three motions of the photon? In last section, we start from RRF to build timeline equations
(\ref{tauMotion})-(\ref{rhoMotion}) and then build equation of motions. In RRF, the timeline equations of $\eta$ and $\rho$ are in $t_1-t_2$ plane, 
the motion of $\eta$ and $\rho$ does not contain $t_0$. Then in observer's RF, the change of x on the path of $\eta$ and $\rho$ contains $t_0$ 
due to Lorentz transformation. But for photon, there is no 
RRF. Also the wave speed and group speed are the same for photon.  We can start building photon's equation of motion from an observer’s view. 
For a photon propagates at z direction, the timeline equation is.
\begin{equation}
\vv{\tau}  = t_0 \vv{q_0} + f(t_r)(cos(\omega t_0) \vv{q_1} - isin(\omega t_0)) \vv{q_2})
\label{tauMotion3}
\end{equation}  
Rewrite above equation as $t_0, t_1, t_2$ component: 
\begin{eqnarray}
t_1 = cos( \omega t_0), \; \;  t_1 = i sin( \omega t_0)
\label{tauMotion4}
\end{eqnarray}
The equation of motion of $\tau$ is
\begin{equation}
\vv{x}(\tau)  = ct_0 \vv{e_z} + F(r) e^{-i\omega t_0} \vv{e_\perp}
\label{equationMotionPhoton}
\end{equation}  
\begin{equation}
\vv{e_\perp} = \vv{e_x} \pm i \vv{e_y}
\label{timeAngleMapping1}
\end{equation}  
where F(r) is a function of the distance from x to the center of the motion,
"+" sign is for the right-hand circular photon, "-" sign is for the left-hand circular photon.
Different from 0-spin particle, the rotation of $\tau$ is associated with space rotation motion, $F(r)(e^{-i\omega t_0} \vv{e_\perp}$.
The imaginary number i in (\ref{timeAngleMapping1}) shows that it takes 1/4 of period to rotate from x axis to y axis. 
To understand (\ref{timeAngleMapping1}), we can substitue it into (\ref{equationMotionPhoton}), then the real part of the function becomes  
\begin{equation}
Re(\vv{x}(\tau))  = ct_0 \vv{e_z} + F(r)(cos(\omega t_0) \vv{e_x} \pm  sin(\omega t_0) \vv{e_y})
\label{equationMotionPhotonReal}
\end{equation}  
It's a rotation in space. So we have a 1-1 mapping between time angle of a timeline vector on $t_1-t_2$ plane $t_\theta$ to the angle of 
a space vector on x-y plane $\theta$: 
\begin{equation}
\theta = t_\theta  
\label{spinAngle}
\end{equation}
The timeline equation of $\eta$ and $\rho$ are:
\begin{equation}
\vv{\eta}  = \eta (cos(kz) \vv{q_1} + isin(kz) \vv{q_2})
\label{etaMotion3}
\end{equation}  
\begin{equation}
\vv{\rho}  = i\rho (cos(kz) \vv{q_1} - sin(kz) \vv{q_2})
\label{rhoMotion3}
\end{equation} 
The equation of motion of $\eta$ and $\rho$ are:
\begin{align}
\vv{x}(\eta)  = z \vv{e_z} + \eta e^{ikz} \vv{e_\perp} \; , \; \; \; \; \; \;  z = c \eta
\label{equationMotionPhotonEta}
\end{align}  
\begin{align}
\vv{x}(\rho)  = z \vv{e_z} + i\rho e^{ikz} \vv{e_\perp} \; , \; \; \; \; \; \; z = c \rho
\label{equationMotionPhotonRho}
\end{align}  
where $\vv{e_\perp}$ is defined in (\ref{timeAngleMapping1}). Equations (\ref{equationMotionPhotonEta}) (\ref{equationMotionPhotonRho}) show that 
when $t_0$ is unchanged, the velocity of motion $\eta$ ($\rho$) has three components: one component points to z direction with constant velocity, 
one component on x-y plane pointing away to the center of the motion, one component rotates around z. Note: without change of $t_0$, $\eta$
($\rho$) can only travel limited distance due to the boundary condition of $t_1-t_2$ plane. The particle will repeat a time loop with $\eta$ 
($\rho$) motion within the distance. 

Let magnitude of inverse velocity of $\eta$ equals electric field $\vv{E}$, magnitude of inverse velocity of $\rho$ equals magnetic field $\vv{B}$,
and wave-function of $\eta$ ($\rho$) is the combinations of $\eta$ ($\rho$) and $\tau$, then we have 
\begin{eqnarray}
\psi_E = R_1(E_x \vv{e_x} \pm i E_y \vv{e_y}) e^{i(kz - \omega t)} \\
\psi_B = R_2(iB_x \vv{e_x} \mp B_y \vv{e_y}) e^{i(kz - \omega t )}
\label{wavefunctionPhoton}
\end{eqnarray}
where $R_1, R_2$  are constants. This is the wave function of the photon. The probability of finding the photon is: 
\begin{equation}
P = |R_1R_2 \vv{E} \times \vv{B}|
\label{probabilityPhoton}
\end{equation}
 The above equations are for circular polarized photon. The linear polarized photon is when left-hand circular timeline vectors of a photon
crossing to its right-hand circular timelines, by using vector addition, it will create a linear polarized photon.
 
\subsection{What is spin?} \label{SPIN}

In section II, motion $\tau$ only has rotation in time coordinate, no space rotation maps to the motion. For photon, equations 
(\ref{equationMotionPhoton}) (\ref{equationMotionPhotonEta}) (\ref{equationMotionPhotonRho}) show the mapping between rotation of time and rotation of space.
For photon, the 1-1 mapping is described in  (\ref{spinAngle}), that is how photons obtain spin.  In general, 
for n-spin particle, the mapping would be $t_\theta = n \theta $. Here is the definition of what spin is: 
Spin of a particle is a rotating timeline vector of a free particle creates a circular motion in space. 

In classical physics, a circular motion has centripetal acceleration. That is because the secondary derivative of $t_0$ is non-zero. In equations
(\ref{equationMotionPhoton}) (\ref{equationMotionPhotonEta}) (\ref{equationMotionPhotonRho}), timeline on $t_1-t_2$ plane is circular, 
the path in space is also circular. It is basically a circle line divided by a circle. The particle's space coordinate on each axis is divided by 
the coordinate on each time axis, the derivative is constant, the secondary derivative on $t_0, t_1,t_2$ are zero, no acceleration! 
When the mapping between timeline and space line is a conformal map, the acceleration can be zero.

\subsection{Derive Maxwell's equations from quaternion Cauchy-Riemann equations} \label{MaxwellEq}

Motions $\tau$, $\eta$, $\rho$ are three independent motions of a particle. Since the motions are from the same particle, there must be relationships between those motions. For 
photon, that is Maxwell's equations. In relativity, the square of length of time is negative, $x_0 =it$, where i is the imaginary number. We also use imaginary number i when 
we map the rotation of time to rotation of space in section \ref{PHOTON}.  In general, when we have 3-d time, we could 
need 3 different imaginary numbers. We can use quaternions I,J,K, where $I^2 = J^2 = K^2 = -1$. Another reason to use quaternion is that 
velocity is defined by space vector
divided by time. In 3-d time, time is a vector. We cannot have a number or a vector divided by a vector, but a number can be divided by a quaternion. 

The typical quaternion can be written as:
\begin{equation}
q = t + Ix + Jy + Kz
\label{quaternion1}
\end{equation}
In 1935, FUETER \cite{FUETER}\cite{Sudbery} extended Cauchy-Riemann equations for a complex function f to quaternions: if f to be regular, it satisfied the equation
\begin{equation}
\partial_t f + I\partial_x f + J \partial_y f + K \partial_z f = 0
\label{Cauchy}
\end{equation} 
We also know that the Pauli matrices equal to imaginary quantities I,J,K by the imaginary number i:
\begin{eqnarray}
 I = -i\sigma_{x} \;\;\;
 J = -i\sigma_{y} \;\;\;
 K = -i\sigma_{z} \;\;\; 
\label{pauli1}
\end{eqnarray}  
In 3-d time coordinate, $\vv{E}$ and $\vv{B}$ are on the same plane, they are perpendicular to each other: $E = iB$. We can use J. Edmonds 
notation \cite{Edmonds}:
\begin{equation}
F = (E^k\sigma_k - icB^k\sigma_k)
\label{F1}
\end{equation}
\begin{equation}
P = i\hbar \sigma_\mu \partial^{\mu} \equiv i\hbar (\sigma_0 \partial^0 - \sigma_1 \partial^1 - \sigma_2 \partial^2 - \sigma_3 \partial^3)
\label{P1}
\end{equation}
where c is speed of light, sum over repeated indices. 
Substitute (\ref{F1})(\ref{pauli1}) into (\ref{Cauchy}), we get the equation
\begin{equation}
PF = 0
\label{equationEdmonds2}
\end{equation}
This is Maxwell's equation in term of quaternion. By simply multiply out all terms in (\ref{equationEdmonds2}), then equate coefficients on both sides for terms $\sigma_{\mu}$, we 
will get the regular Maxwell in vacuum. We see that as long as we know E and B are two motions perpendicular to each other in $t_1-t_2$ plane,  Maxwell’s 
equations in vacuum is the Cauchy-Riemann equations in 3-d time physics .     

Equation (\ref{Cauchy}) only works if f is regular. When there is source for electromagnetic field, we can treat source as discontinuities of the space-time, the equation 
(\ref{equationEdmonds2}) can be extended as \cite{Edmonds}:
\begin{eqnarray}
PF = \epsilon J 
\label{Edmonds3}
\end{eqnarray}  
where $\epsilon$ is a constant. $J = qc\sigma_0 + j^k\sigma_k$ is the source, where q is the charge density and $j^k$ is the current density. Both equations (\ref{equationEdmonds2})
(\ref{Edmonds3}) are derived by J. Edmonds in 1977 \cite{Edmonds}, here we give the 3-d time meaning to the equations. 

We can also derive Maxwell’s equations using 6-dimensional Einstein field equations. We will discuss that in unification of gravitational and electromagnetic fields. 

\subsection{Equations of motion for 1/2-spin particle } \label{halfSpinEquations}

Equations (\ref{equationEdmonds2}) contain 6 dimensions. The quaternion provides 4 dimensions represents (t,x,y,z). E (B) represent a two-dimensional vector on $(t_1,t_2)$ plane, 
each has 2 dimensions -- real part and imaginary part. 
Quantum physics is to use 4 measurable dimensional (t,x,y,z) as variables to study the two unknown timelines $\eta$, $\rho$.
 Each timeline has two components ($t_1,t_2$ directions). The overall equations should be quaternions act on a function having two components, this is how we are going to 
 struct the quaternion form of Dirac equation. 
 
Let's write wave-function as quaternion
\begin{equation}
F = \psi = \frac{1}{m_0}(p^0 - \sigma_k p^k) \phi
\label{WeylWave0}
\end{equation}
where $p^0 = E/c$, E is the energy of the particle. R is normalization constant. Then use equation (\ref{Edmonds3}), we have 
\begin{eqnarray}
PF = m_0 \phi 
\label{Edmonds4}
\end{eqnarray}  
where $m_0 \phi$ serves as the source. This is the Dirac equation of positive energy in quaternion form . We will explore more detail of the reason behind (\ref{Edmonds4}).  
 
Now we build timeline equations of $\eta$,$\rho$ for 1/2 massive particle. 
When we discuss the 3-d time geometry of 0-spin and 1-spin particle, it has cylindrical symmetry. The base of the cylinder is a plane. For 1/2 massive particle, we assume that 
the base of the cylinder is spherical. 

\begin{figure}
  \includegraphics[scale=1.0]{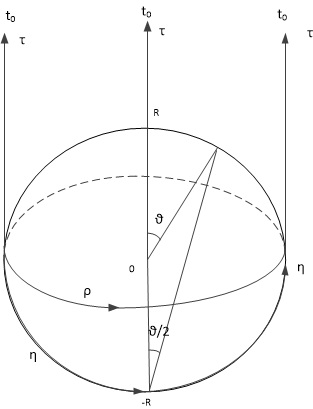}
  \caption{The time geometry of 1/2 spin particle. The base of the cyliner is a hemisphere. Timeline $\eta$ is along longitude line, timeline $\rho$ is along latitude line, 
  $\tau$ is perpendicular line. }
  \label{fig:geometryOfhalfSpin}
\end{figure}

Fig \ref{fig:geometryOfhalfSpin} shows time geometry of 1/2 spin particle. The base of the cylinder is a time sphere with radius R. The center of the base is at the south pole of the sphere 
with coordinates $t_0 = -R, t_1 = t_2 = 0$; timeline $\eta$ along the longitude line, 
timeline $\rho$ is the latitude of the sphere perpendicular to $\eta$.  $\tau$ is the same as 0-spin particle. From geometry in Fig \ref{fig:geometryOfhalfSpin}, timeline equations for $\eta$ are:
\begin{equation}
t_0 = R cos(\frac{\theta}{2}) \; \; , \; \; t_r = sin(\frac{\theta}{2})
\label{halfspinEta}
\end{equation} 
where R is constant. For $\rho$:
\begin{equation}
t_1 = t_r cos(\phi) \; \;, \; \; t_2 = it_r sin(\phi)
\label{halfspinRho}
\end{equation}
Substitute the second function of (\ref{halfspinEta}) into (\ref{halfspinRho}), then combine $\eta$ and $\rho$, we get a 2-components wave function
\begin{equation}
\psi = R \left(\begin{array}{cc}
	cos(\frac{\theta}{2}) \\
	sin(\frac{\theta}{2}) e^{i\phi} 
   \end{array}\right)
\label{SpinorWaveFun1}
\end{equation}
The top component is the $t_0$ component of $\eta$, the bottom component is the $t_r$ componet of $\eta$ with the direction rotated with $\rho$. 
This wave-function does not contain space coordinates. 
After removing the constant R,
the equation (\ref{SpinorWaveFun1}) is an eigenfunction of 
\begin{align}
\sigma \cdot n =  \left(\begin{array}{cc}
	cos(\theta) \; \; \; \; &sin(\theta) e^{-i\phi} \\
	sin(\theta) e^{i\phi} \; \; \; \;  &-cos(\theta)
   \end{array}\right)
\label{SpinorWaveMatrix}
\end{align} 
with the eigenvalue +1. 
The wave-function (\ref{SpinorWaveFun1}) is the combination of the direction of timeline vectors $\eta$, $\rho$.
In our previous discussion, for 0-spin and 1-spin particle, wave-funciton is a time vector on $t_1-t_2$ plane, where the phase of 
the function is the direction angle of the vector. Equation (\ref{SpinorWaveFun1}) shows that for 1/2 spin particle, the wave-function vector is not limited 
to $t_1-t_2$ plane anymore, it has three-dimensional directions in time coordinates. 

Now we need to find the mapping from the above timelines to space lines. We can make 1-1 same shape mapping from time sphere to space sphere. 
Since it's the same shape mapping, the magnitude of the inverse velocity should be constant, so we only need to focus on the directions of space and time vector. 
For right-hand spin, the mapping between timeline vector to vector on space sphere can be written as:
 $R(cos\theta \vv{e_z} + sin(\theta) (cos\phi\vv{e_x} + isin\phi\vv{e_y})) $, 
where i represent the 1/4 period phase difference between $e_x$ and $e_y$ when under the rotation of $\rho$. 
$\eta$ motion provides the term related to $\theta$, $\rho$ provides the term related to $\phi$. we can write it as two component wave-function vector: 
\begin{equation}
\vv{\psi} = \left(\begin{array}{cc}
	cos\theta \vv{e_z}\\
	sin(\theta) (cos\phi\vv{e_x} + isin\phi\vv{e_y})
   \end{array}\right)
\label{SpinorSpaceRight}
\end{equation}
Compared to (\ref{SpinorWaveFun1}), (\ref{SpinorSpaceRight}) is a wave-function in space. In previous section, wave-function is the combination of two motions 
$\eta$-$\tau$ and $\rho$-$\tau$ respectively. (\ref{SpinorSpaceRight}) is the combination of motions $\eta$-$\rho$. For 1/2 spin particle, the wave-function is the combination of 
all three motions. For 0-spin particle, the wave-function is inner product of $\vv{e_\eta}$ and $\vv{e_\tau}$  (equation \ref{waveFunction1}) because only the momentum $\vv{p}$ direction 
of $\eta$ and $\rho$ contributes to wave-funciton.  Thus, the final wave-function is:
\begin{align}
\psi & = \vv{\psi} \cdot \vv{e_p} e^{\frac{-i}{\hbar} (Et - px) } 
	 & = e^{\frac{-i}{\hbar} (Et - px) } \left(\begin{array}{cc}
	cos\theta \\
	sin(\theta) (cos\phi - isin\phi)
   \end{array}\right)
\label{SpinorSpaceRight1}
\end{align}
where $cos\theta = e_z \cdot e_p, sin\theta cos\phi = e_x \cdot e_p, sin\theta  sin\phi = e_y \cdot e_p $.
For left-hand spin, the equation can be written as  
\begin{equation}
\psi = e^{\frac{-i}{\hbar} (Et - px) } \left(\begin{array}{cc}
	-cos\theta \\
	sin(\theta) (cos\phi + isin\phi)
   \end{array}\right)
\label{SpinorSpaceLeft1}
\end{equation}

To make the above function look like Dirac wave-function, we inroduce $\chi$
\begin{equation}
\chi = \left(\begin{array}{cc}
	c_1\\
	c_2
   \end{array}\right)
\label{chi1}
\end{equation}
where $c_1 = 1, c_2 = 0$ for right-hand spin (+ spin), $c_1 = 0, c_2 = 1$ for left-hand spin (- spin).
Writing (\ref{SpinorSpaceRight1}) (\ref{SpinorSpaceLeft1}) together and switch the top and bottom of (\ref{SpinorSpaceLeft1}), we have 
\begin{equation}
\psi = e^{\frac{-i}{\hbar} (Et - px) } (\sigma \cdot \vv{e_p}) \chi
\label{SpinorSpacePauli}
\end{equation} 

The above equation is a space wave-function . Dirac wave-function also has $t_0$ component. For the single-color plane wave, 
the $t_0$ component should only be $ a e^{\frac{-i}{\hbar} (Et - px) }$, a is a constant. We can write time component on the top of two components wave-function (\ref{SpinorSpacePauli})
to form a 4-components wave-function, but what is the ratio between $t_0$ component and space component? From Fig \ref{fig:geometryOfhalfSpin}, if we map the $t_1 -t_2$ plane to space, due to the 
opposite sign between the $dt^2$ and $dx^2$, $cos\frac{\theta}{2}, sin\frac{\theta}{2}$ becomes $cosh\frac{\theta}{2}, sinh\frac{\theta}{2}$ (Lorentz transformation), then the ratio of $t_0$ component
to space component is $coth\frac{\theta}{2}$. The 4 components wave-function can be written as:
 \begin{equation}
\psi = R\left(\begin{array}{cc}
	cosh\frac{\theta}{2} \\
	0 \\
	sinh\frac{\theta}{2} (\sigma \cdot e_p ) \chi
   \end{array}\right) e^{\frac{-i}{\hbar} (Et - px) } 
\label{DiracWaveFun}
\end{equation}
where $cosh\frac{\theta}{2} = \sqrt {\frac{\gamma + 1}{2}}, sinh\frac{\theta}{2} = \sqrt {\frac{\gamma + 1}{2}}, \gamma = \frac{1}{\sqrt{1-\beta^2}}, \beta = \frac{v}{c}$. Using 
both $E = \gamma m_0$ and $ p = \gamma m_0v$, and add normalization constant, we get 
\begin{equation} 
\psi = \sqrt{\frac{1}{2m_0(E + m_0)}} \left(\begin{array}{cc}
	E + m_0 \\
	0 \\
	(\sigma \cdot \vv{p} ) \chi
   \end{array}\right) e^{\frac{-i}{\hbar} (Et - px) } 
\label{DiracWaveFun1}
\end{equation}
This is Dirac wave-function. 

\subsection{Double coverage and Pauli exclusion principle } \label{DoubleCoverage}

There is one issue with spherical based cylinder. When projecting the whole sphere to $t_1-t_2$ plane, it double covers $t_1-t_2$ plane. Therefore, for a single 1/2 spin particle,
its base can only be a half sphere: either upper hemisphere or lower hemisphere, the center of the base would be either at the north pole or south pole. The equator is 
the edge of the $t_1-t_2$ plane. 

Let's look at the difference among 0-spin (scalar) wave-function, 1-spin (vector) wave-function and Dirac spinor wave-functions. 
For scalar wave-function,  Two motions $\eta$ ($\rho$)  and $\tau$ formed wave-function $\psi$ ($\psi^*$). The two motions are identical. Timelines for 
both motions are within $t_1-t_2$ plane.
For vector wave-function, $\eta$, $\rho$ each forms wave-function with $\tau$, the two motions are not identical, such as $\vv{B}$ and $\vv{E}$. Timelines for 
both motions are within $t_1-t_2$ plane.
For spinor wave-function, all three motions: 
$\eta$, $\rho$, $\tau$ together form the wave-function. The direction of timeline $\eta$ is three-dimensional instead of limited within $t_1-t_2$ plane.  
There are two rotations in time coordinates: one from $\tau$, one from $\rho$. 
The Dirac wave-function shows the full picture of the three worldlines. 

When we discuss Bose-Einstein condensation, we can have many particles at the same space location at the same $t_0$ as long as they have a phase difference, 
 none of them would collide to others because they don't meet at same 3+3 space-time coordinate point. With spherical geometry, we cannot have two particles at the 
same half of hemisphere because no matter how 
much phase difference at latitude or longitude, the two particles would eventually collide to each other at south pole (north pole). There are at most two 1/2 spin particles
can stay at the same space location at the same $t_0$, one particle at south half of hemisphere, the other at the north half of hemisphere. They need to always move away from 
each other, one spin towards +z direction, the other spin towards -z direction. This is the physics behind Pauli exclusion principle. 

If we change t to -t, $\partial_t$ to $-\partial_{t}$, we will get the negative energy solution of Dirac equation. One question puzzled physicist is that why the particle 
doesn't jump to negative energy level? The popular answer is that all negative energy level are filled by particles, that's so called the negative energy sea. From the view 
of 3-d time physics, the negative energy means timeline $\tau$ points to negative $t_0$ direction, since all timelines are directed, $\tau$ always point to positive $t_0$, the particle 
cannot jump to negative energy level.

\subsection {6-dimensional Einstein field equations for 1/2 spin particle } \label{halfSpinFieldEquation}

The metric of 1/2 particle can be written as :
\begin{align}
\left( \hat{g}_{AB} \right) &=  \begin{bmatrix}
   g_{\alpha\beta}   &   0  &   0 \\
   0  &   -\psi^2   &  M \psi^2	\\
   0  &    M \psi^2     & 1-M^2 \psi^2 		
   \end{bmatrix}  
\label{6dMetric_half}
\end{align}
where 
\begin{equation}
\psi = R e^{\frac{-i}{\hbar}(Et_0 - px)}e^{it_\phi}   
\label{localInertialHalf}
\end{equation}
R is the radius of the sphere, which is constant. $x_0 \equiv t_0$. $x_4 \equiv t_\phi $ is time angle in $t_1-t_2$ plane.
The $e^{it_\phi}$ term comes from the rotation of the $\rho$ timeline.   
$\partial_5 \psi = 0$,  $\partial_4 \psi = i\psi$.  M is a scalar field acting on $x_5$. We choose $c=1$.  

The interval can be written as 
\begin{equation}
ds^2 = g_{\alpha\beta}dx^\alpha dx^\beta + dx^5 dx^5 - (\psi dx^4 - M \psi dx^5)^2  
\label{interval_halfspin}
\end{equation}

Let 
\begin{equation}
dx^{4'} = dx^4 - Mdx^5 , \; \; dx^{\alpha '} = dx^{\alpha},  \; \; dx^{5'} = dx^{5} 
\label{localInertialHalf1}
\end{equation}
using $\partial_A dx^A = \delta_{AB}$, we have
\begin{equation}
\partial_{\alpha} ' = \partial_{\alpha} ,\; \; \partial_{4}' = \partial_{4}  \;, \; \partial_{5}' = \partial_{5} + M\partial_4 
\label{localInertialHalf2}
\end{equation}
Using equations (\ref{interval_halfspin}), (\ref{6dChristRicci}), and let the Ricci tensor $R_{44} = 0$, we get:
\begin{equation}
{\partial^{\alpha}} ' \partial_{\alpha} '  \psi + {\partial^5} ' \partial_5 ' \psi = 0  
\label{6dSETensoRhalf}
\end{equation}
where 
\begin{equation}
\partial_5 ' \psi \equiv (\partial_5 + M \partial_4) \psi = i M \psi  
\label{halfspin_partial_5}
\end{equation}
If we let $M =\frac{m_0}{\hbar}$, where $m_0$ is the particle's rest mass. Then equation  (\ref{6dSETensoRhalf}) becomes Klein-Gordon equation. 
We see 1/2 spin particle obtains its mass through a scalar field M. i.e. A scalar field is source of rest mass of 1/2 spin particle. The $\alpha \beta-$, 
$\alpha 5- $, 55-, $\alpha 4-$components of Einstein 
equations (\ref{5dEFE1}) become:
\begin{equation}
 \frac{1}{\phi}\partial_{\alpha}\partial_{\beta}\psi = -\kappa \hat{T}_{\alpha\beta} 
\label{6dMSTensor_01}
\end{equation}
\begin{eqnarray}
\frac{-im_0}{\hbar \psi} \partial_{\alpha} \psi = \kappa \hat{T}_{5\alpha} = \kappa \hat{T}_{\alpha 5} \\
\label{6dMSTensor_11}
\frac{m_0 ^2}{\hbar ^2} = \kappa \hat{T}_{55}     \\
\label{5dMSTensor_21}
 \kappa \hat{T}_{4 \beta} = \kappa \hat{T}_{\alpha 4} = 0 
\label{6dSETensor_31}
\end{eqnarray}

The interval $ds^2$ is the square of length. It cannot be used to include spinor. To describe
spinor, let : 
\begin{align}
dx^1 dx^1 & = (iI {dx^1}') (iI {dx^1}')  \\ 
dx^2 dx^2 & = (iJ {dx^2}') (iJ {dx^2}')  \\  
dx^3 dx^3 & = (iK {dx^3}') (iK {dx^3}') 
\label{quaternion_half}
\end{align}
 where I, J, K are quaternions. Let $\sigma^1 = iI, \sigma^2 = iJ, \sigma^3 = iK$. To mimic Maxwell equations $B = \nabla \times A$ and 
 $\nabla \times B = - \frac{\partial E}{\partial t}$, we definte a spinor $\psi_q$:
\begin{equation}
\psi_q = i\hbar (\partial_t + \sigma^k \partial_k) \psi 
\label{new_spinor1}
\end{equation}
and a scalar $\psi_s$
\begin{equation}
\psi_s = i\hbar M \partial_4 \psi
\label{new_spinor2}
\end{equation}
where k=1,2,3. Then (\ref{6dSETensoRhalf}) can be written as 
\begin{equation}
\sigma_\mu \partial^{\mu} \psi_q = - M \partial_4 \psi_s 
\label{newDirac0}
\end{equation}

The definition of $\psi$ (\ref{localInertialHalf}) is a time vector on $t_1-t_2$ plane. We know that time vector 
of 1/2 particle has three dimensions. In section \ref{halfSpinEquations}, we use two componet vector to represent a time vector's 3-dimensional direction.  
Thus, we should redefine $\psi$ in (\ref{localInertialHalf}):
\begin{equation}
\psi -> \hat{\psi} = R \left(\begin{array}{cc}
	\psi_1 \\
	\psi_2
   \end{array}\right) e^{\frac{-i}{\hbar}(Et_0 - px)}e^{it_\phi}   
\label{diracHatlHalf}
\end{equation}
where $\psi_1^2 + \psi_2^2 = 1$. Substitute $\hat{\psi}$ in (\ref{new_spinor1}) (\ref{new_spinor2}), (\ref{newDirac0}) and 
let $\sigma^k$ as 2X2 Pauli matrices. Then 
\begin{equation}
\sigma_\mu \partial^{\mu} \hat{\psi}_q = -M \partial_4 \hat{\psi}_s 
\label{newDirac}
\end{equation}

Again we get Dirac equation in quaternion form for the 1/2 spin particle with positive energy. But if we write $\psi$ as a $1 \times 2$ matrix, and 
$dx^k dx^k -> \sigma^k dx^k \sigma^k dx^k $, does it  
make metric (\ref{6dMetric_half}) 12 dimensions? The answer is no. Einstein field equation is based on the least action 
principle on a single worldline. The path of the particle in Einstein field is geodesic. Now with three time dimensions, the particle  
has three worldline motions, theoretically Einstein field equation cannot accurately describe the particle's motion. To use Einstein field 
equation, we need to combine the motions and treat the motions together as a group. The definition (\ref{localInertialHalf}) is a wave-function,
which contains two motions. The two components complex function (\ref{diracHatlHalf}) represents 
a time vector with three dimensional direction, which includes three motions. Therefore, the matrices of $\psi$ and $\sigma$ are not extra space-time dimension, 
they come from the motions driven by two extra proper time vectors,  metric (\ref{6dMetric_half}) is still a metric of 3+3 spacetime. 
We don't open up the matrices calculation when we deduct the Ricci tensor and Riemann tensor. Thus, when we replace $\psi$ and $\partial_k$
with $\hat{\psi}$ and $\sigma^k \partial_k$, most part of the Einstein equations (\ref{5dEFE1}) keep the same. But because of 
$\sigma_i \sigma_j = - \sigma_j \sigma_i$, then (\ref{5dMSTensor_1}) changes to  
\begin{equation}
\sigma^{\alpha}\frac{-im_0}{\hbar R^2} \hat{\psi}^+ \partial_{\alpha} \hat{\psi} = \kappa \hat{T}_{5\alpha} = - \kappa \hat{T}_{\alpha 5} 
\label{6dMSTensor_12}
\end{equation}
where $\hat{\psi}^{+}$ is the complex conjugate of $\hat{\psi}$.

How do we interpret (\ref{newDirac})? In classical physics, the velocity is $\frac{dx}{dt}$, where x is 3-d space vector, t is scalar. 
In 3-d time, both time and space are vectors. We want to get inverse velocity: timeline vector/ 3-d space vector. To divide by a vector, we 
can use quaternion. Define quaternion inverse velocity operator: 
\begin{equation}
\hat{u} = \frac{\partial}{\partial_q} \equiv \sigma^k\partial_k  
\label{inversionVelocityV0}
\end{equation}
where k=1,2,3. Since metric has the unit of square of length, $\psi$ has the unit of time length. then $\hat{u}\psi$ is inverse velocity, 
and $\hat{u} \hat{u}\psi$ is inverse acceleration. Equation below is an oscillation equation:
\begin{equation}
\hat{u} \hat{u} \psi = (- \omega^2 + \omega_0 ^2) \psi    
\label{inversionVelocityV}
\end{equation}
Let $\omega = \frac{E}{\hbar}, \omega_0 = \frac{m_0}{\hbar} $. Then (\ref{inversionVelocityV}) becomes (\ref{newDirac}). Compared to classical oscillation equation 
\begin{equation}
\ddot{x} = -\omega^2 x     
\label{oscallation1}
\end{equation}
Equation (\ref{inversionVelocityV}) clearly shows that $\psi$ is a time vector, $\hat{u} \hat{u} \psi$ is the inverse acceleration, energy E causes the oscillation of 
time vector, which explains why the whole particle's time coordinates rotates with $t_0$, $m_0$ is a counter oscillation scalar 
potential. That's the reason we see M acting as a separate scalar field in metric (\ref{6dMetric_half}), and acting as source in (\ref{Edmonds4}).

\section{Interactions and Unified theory} \label{INTERACTION}

\subsection{Distorted space-time geometry and interactions} \label{DistoredInteractions}

In general relativity, the gravitational interaction acts on an object through distorted space-time. In this paper, the energy of the particle creates the distorted 3-d time geometry,
which leads to the motion of the 0-spin, 1-spin and 1/2-spin particles. In other words, the fields described in quantum field theory is a distorted 3-d time + 3-d space. The 
distorted space-time governs the particle's motion even without interaction. We expanded the principle of general relativity to the free particles. Any interactions 
among the particles would be described as one particle's space-time geometry changing by the other particles' energy.  Then the logic would be: all interactions should be 
unified by equations of space-time geometry. Can we expand gravitational Einstein field equations to other interactions? There is one problem: 

Einstein field equations is based on the principle of least action on an object's worldline. But in 3-d time physics, the particle does not travel through a 
single worldline. Instead, the paths are 3-dimensional worldlines  (3-dimensional world fields). With 3-d time motions, geodesic path on each worldline may not 
always be the least action for the particle as a whole. We might need to find the minimum surface or minimum volume in the space-time geometry. A interesting
observation is that the wave-function of 0-spin particle is a helicoid formed by motions $\tau$ and $\eta$($\rho$), and helicoid is known as a complete embedded 
minimal surface of finite topology with infinite curvature. If the physical system has certain space-time symmetry, we can treat the wave-function as one classic
 object motion so that we can still apply Einstein field equations on the 3-d time motions. That's how we derive the field equation on 0-spin free particle from 6-dimensional 
Einstein field equations.

We are going to use 6-dimensional Einstein fields equations to unify gravitational interaction and electromagnetic interaction. Then we will discuss strong interaction.

\subsection{What is gauge transformation invariance?} \label{GaugeTransformation}

First let's define the difference between fields and particles. The field is distorted 3+3 dimensional space-time. The 3-d time fields are directed 3-d
time flow. A particle is a sizeless energy point. The particle moves along 3-d space driven by 3-d timelines. The energy of particle will cause 
the particle's time field rotation with worldline $\tau$. When a particle doesn't have $\tau$ motion, it is a field.   
 
Now let's understand what the gauge field is. For photon, the gauge field is represented by a choice of electromagnetic 4-vector $(\phi, \vv{A})$. 
The space components of gauge field vector $A_\alpha$ 
is defined as $\vv{B} = \nabla \times \vv{A}$. When we take the derivative the motion of 1-spin particle (photon), the electron magnetic fields $\vv{E}$ and $\vv{B}$ are 
wave-functions. In 3-d time physics, this means  $\vv{E}$ and $\vv{B}$ are inverse velocities . Since $\vv{B}$ has the form of  $\vv{A}$ divided by x,
$\vv{A}$  is a time flow vector. $\vv{A}$ will act as part of space-time metric. In classical electromagnetic theory, 
vector $A_\alpha$ does not represent a real physical property. But Aharnove-Bohm effect
showed that $\vv{A}$ does have observable physical effect. From the above discussion, we see that in Aharnove-Bohm experiment, even though there are no electron 
magnetic fields $\vv{E}$ and $\vv{B}$ outside the confined area, the vector $\vv{A}$ still changes the metric of time outside the area. Therefore, it changes 
the direction of time vector of the electron. i.e. $\vv{A}$ changes the geometry of time angle $t_\phi$ in $t_1-t_2$ plane. 

In quantum gauge field theory, the interaction can be derived by gauge transformation invariance property of the particle's Lagrangian. For a quantum field 
$\psi$, the global gauge invariance is that the system doesn't change if we replace $\psi$ by $\psi e^{i\alpha}$. From 3-d time physics, we know that the 
phase of a quantum field is just the direction of time vector on $t_1-t_2$ plane. We can choose any random direction as start angle $\phi = 0$, the system 
doesn't change. Therefore, we have global gauge invariance. When time metric contains gauge field $\vv{A}$, the metric of time angle becomes a function of x, 
 the system shouldn't depend on the choice of initial angle $\phi$ even thought $\phi$ changing over x, therefore, we need to have local gauge invariance.
The system should be unchanged under the transformation of  $\psi -> \psi e^{\alpha (x)}$. 

\subsection{Photon in gravitational field} \label{PhotonGravitional}

It is trivial to extend Kaluza-Klein equations to 6 dimensions on photon.  The 6 dimensional metric can be written as:
\begin{align}
\left( \hat{g}_{AB} \right)  &= \begin{bmatrix}
   g_{\alpha\beta} - \kappa^2 A_{\alpha} A_{\beta}  &   -\kappa A_{\alpha}  &  0 \\
   - \kappa A_{\beta}  &   -1  &  0 \\
    0  &  0 & -1 
   \end{bmatrix}
\label{6dMetric1S}
\end{align}
For massless particle, we assume $\partial_5 A_\alpha = 0$. We also assume $\partial_4 A_\alpha =0$. The $\alpha \beta-$, $\alpha 4- $, 
and 44-components of equations (\ref{5dEFE1}) become:
\begin{eqnarray}
G_{\alpha\beta} = \frac{\kappa^2}{2} 
   T_{\alpha\beta}^{EM} \;, \; \;  
\nabla^{\alpha} \, F_{\alpha\beta} = 0 \; \; \; , \; \; \;
F_{\alpha\beta} F^{\alpha\beta} = 0 \; \; \; 
\label{6dFieldEquns1S}
\end{eqnarray}
The second of above equations is Maxwell equation. The third of equation is true for 
plane electromagnetic wave-function which 
is the case of single free-photon. The first equation describes the gravitational force acting on photon, where 
$G_{\alpha\beta} \equiv R_{\alpha\beta} - RG_{\alpha\beta}/2$ is the Einstein tensor. $T_{\alpha\beta}^{EM} \equiv g_{\alpha\beta} F_{\gamma\delta} F^{\gamma\delta}/4 -F^\gamma _\alpha F_{\beta\gamma}$ 
 is the electromagnetic energy-momentum tensor, and $F_{\alpha\beta} \equiv \partial_\alpha A_\beta - \partial_\beta A_\alpha$. 
The scaling parameter $\kappa \equiv 4\sqrt{\pi G}$, where G is gravitational constant.   

Equations (\ref{6dFieldEquns1S}) describes a free photon moving in a gravitational fields. To include electromagnetic interaction in 6-d Kaluza–Klein equations, we 
need to add massive charged particle terms in the metric.

\subsection{The electromagnetic field and integer spin charged particle interaction using Einstein field equations} \label{EFIntegerSpin}

Considering a particle with a cylindrical symmetry in 3-d time. Timeline $\eta$ moves along the radius in $t_1-t_2$ plane timeline, $\rho$ moves along the angular 
direction. Such particle could be a particle with integer spin or spinless particle depending on if the changing of time angle mapping to the changing of space angle. 
The metric of such particle can be written as: 
\begin{equation}
ds^2 = - dt_0^2 - dt_r^2 - (f(t_r) e^{\frac{-i}{\hbar}(Et_0 - px)}e^{it_\theta})^2   dt_\theta ^2    
\label{cylinder_metric_halfspin}
\end{equation}
Compared to metric of 0-spin particle (\ref{cylindar_metric_0}), the extra term $e^{it_\theta}$ is coming from the circular motion 
of timeline $\rho$. 
Let 
\begin{equation}
\phi = f(t_r) e^{\frac{-i}{\hbar}(Et_0 - px)} e^{it_\theta}   
\label{phi_halfspin}
\end{equation}
And let $f(t_r) = e^{\frac{-m_0}{\hbar}t_r}$, $x_4 = t_\theta$, $x_5 = t_r$. We see
\begin{equation}
\partial_5 \phi = -\frac{m_0}{\hbar} \phi \;,	\;	\;	\; \; \;
\partial_4 \phi = i  
\label{partial_4_5}
\end{equation} 

The 6-dimensional metric of a charged particle can be written as:  
\begin{align}
\left( \hat{g}_{AB} \right) &=   \begin{bmatrix}
   g_{\alpha\beta} - g_e^2 \phi^2 A_{\alpha} A_{\beta}  &  - g_e\phi^2 A_{\alpha}  &   0 \\
    -g_e \phi^2 A_{\beta}   &   -\phi^2   &  0	\\
   0 & 0 &  -1 		
	\end{bmatrix} 
\label{6dMetric_charge}
\end{align}
where $\phi$ is described in (\ref{phi_halfspin}), $g_e$ is the coupling constant, $g_e = \frac{e}{\hbar}$, e is the charge of electron. 

Under this metric, interval becomes:
\begin{equation}
ds^2 = g_{\alpha\beta}dx^\alpha dx^\beta - dx^5 dx^5 - (\phi dx^4 + g_e \phi A_\alpha dx^\alpha)^2  
\label{interval_charged}
\end{equation}
let 
\begin{equation}
dx^{4'} = dx^4 - g_e A_{\alpha} dx^{\alpha}, \; \; dx^{\alpha '} = dx^{\alpha}  \;, \; dx^{5'} = dx^{5} 
\label{localInertial}
\end{equation}
using $\partial_A dx^A = \delta_{AB}$, we have 
\begin{equation}
\partial_{\alpha} ' = \partial_{\alpha} - g_e A_{\alpha} \partial_{4} ,\; \; \partial_{4}' = \partial_{4}  \; ,\; \partial_{5}' = \partial_{5} 
\label{localInertial2}
\end{equation}
Under the new coordination, the interval (\ref{interval_charged}) becomes:
\begin{equation}
ds^2 = g_{\alpha\beta}{dx^\alpha} ' {dx^\beta} ' - {dx^5} ' {dx^5} ' - \phi^2 {dx^4} ' {dx^4 }' 
\label{interval_charged2}
\end{equation}

Using equations (\ref{interval_charged2}), (\ref{6dChristRicci}), and let $R_{44} = 0$, the $\alpha \beta-$, $\alpha 5- $, 55-, 
$\alpha 4- $, and 44-components of Einstein equations (\ref{5dEFE1}) become:
\begin{equation}
 -\frac{1}{\phi}\partial_{\alpha} ' \partial_{\beta} ' \phi = \kappa \hat{T}_{\alpha\beta} 
\label{5dMSTensor_0}
\end{equation}
\begin{eqnarray}
\frac{m_0}{\hbar \phi} \partial_{\alpha} ' \phi = \kappa \hat{T}_{5\alpha} = \kappa \hat{T}_{\alpha 5} \\
\label{5dMSTensor_1}
\frac{m_0 ^2}{\hbar ^2} = -\kappa \hat{T}_{55}     \\
\label{5dMSTensor_2}
 \kappa \hat{T}_{4 \beta} = \kappa \hat{T}_{\alpha 4} = 0 \\
\label{5dSETensor_3}
{\partial^{\alpha}} ' {\partial_{\alpha}} '  \phi - m_0^2\phi = 0  
\label{5dSETensor_4}
\end{eqnarray}

Equation (\ref{5dSETensor_4}) has the format of Klein-Gordon equation except that the derivative
\begin{equation}
\partial_{\alpha} ' = \partial_{\alpha} - g_e A_{\alpha} \partial_{4}  
\label{leastCoupling1}
\end{equation}
Since $\partial_{4} \phi = i$, then we have 
\begin{equation}
\partial_{\alpha} ' \phi = (\partial_{\alpha} - ig_e A_{\alpha}) \phi  
\label{leastCoupling2}
\end{equation} 
This is the minimal coupling equation between the charge and electromagnetic field. Therefore, (\ref{5dSETensor_4}) is the equation for the interaction between 
the particle and electromagnetic field.

Equations (\ref{5dMSTensor_0})-(\ref{5dSETensor_4}) tells us that with the choice of the coordinate, the particle is moving like a free particle in electromagnetic field, this is so called local flat. 
This is analog to an object in free fall elevator in gravitational field. The equation also tells us that the particle carries electromagnetic fields with it,
the external electromagnetic interacts with  particle by increasing $A_{\alpha}$. 

We see there are two conditions that a particle can interact with electromagnetic field. 1) The particle needs to have circular $\rho$ motion so that $\partial_4 != 0$.
2) The particle needs to carry electromagnetic vector $A_\alpha$ in it's metric. 

In above equations, $\phi$ is a scalar field. For vector particle (integer spin), the matric is: 
\begin{equation}
ds^2 = g_{\alpha\beta}dx^\alpha dx^\beta - dx^5 dx^5 - (\phi dx^4 + g_e \phi A_\alpha dx^\alpha + \phi_\alpha dx^\alpha )^2  
\label{interval_charged_vector}
\end{equation}  
and then 
\begin{equation}
dx^{4'} = dx^4 - (g_e A_{\alpha} + \phi_{\alpha}) dx^{\alpha} , \; \; dx^{\alpha '} = dx^{\alpha}  \;, \; dx^{5'} = dx^{5} 
\label{localInertialV}
\end{equation}
\begin{equation}
\partial_{\alpha} ' = \partial_{\alpha} - (g_e A_{\alpha} + \phi_{\alpha}) \partial_{4}, \; \; \partial_{4}' = \partial_{4}  \; , \; \partial_{5}' = \partial_{5} 
\label{localInertial2V}
\end{equation}
where $\phi_{\alpha}$ is the vector components of the integer spin particle. The above equation also includes the interactions between electromagnetic field and 
the vector components of the particle $\phi_\alpha A_\beta$. Substitute (\ref{localInertialV}) (\ref{localInertial2V}) into (\ref{interval_charged_vector}), 
then metric (\ref{interval_charged_vector}) will act like the metric of free spinless particle.

\subsection{The electromagnetic field and 1/2-spin charged particle interaction using Einstein field equations  } \label{EFInteractionHalfSpin}

For charged 1/2-spin particle, the metric is: 
\begin{equation}
\begin{split}
ds^2 &= g_{\alpha\beta}dx^\alpha dx^\beta- dx^5 dx^5 - \\
      &	(\phi dx^4 + g_e \phi A_\alpha dx^\alpha - M\phi dx^5)^2  
\end{split}
\label{interval_charged_halfspin}
\end{equation}
and 
\begin{eqnarray}
dx^{4'} = dx^4 + g_e A_{\alpha} dx^{\alpha} - M dx^5  \\
 dx^{\alpha '} = dx^{\alpha}  \\
 dx^{5'} = dx^{5} 
\label{localInertial2}
\partial_{\alpha} ' = \partial_{\alpha} - g_e A_{\alpha} \partial_{4} \\ 
\partial_{4}' = \partial_{4}  \\
\partial_{5}' = \partial_{5} + M \partial_4
\label{localInertialHalfspin2}
\end{eqnarray}
Then equation (\ref{newDirac}) becomes
\begin{equation}
P'\hat{\psi} ' = -i\hbar M \partial_4 \psi_s  
\label{newDirac_charged}
\end{equation}
where 
\begin{equation}
P' = i\hbar ({\partial^0} ' - \sigma^1 {\partial^1} ' - \sigma^2{ \partial^2} ' - \sigma^3 {\partial^3} ')
\label{P1_dirac}
\end{equation}
\begin{equation}
\hat{\psi} ' = i\hbar (\partial_0 ' + \sigma^k \partial_k ') \phi 
\label{F_dirac} 
\end{equation}
(\ref{newDirac_charged}) is the Dirac equation with the particle - electromagnetic interactions in quaternion form. 
It's easy check that it contains spin-magnetic field interactions. 

\subsection{1-spin massive free meson} \label{oneSpin_1}
 
For 1-spin massive meson, we let 
\begin{equation}
\hat{A}_{\alpha} = A_{\alpha}e^{\frac{-m_0 }{\hbar} x_5}
\label{hatA}
\end{equation}
and
\begin{equation}
\hat{A}_{5} = 0, \;	\;	\;	\;	\; \; \hat{A}_{4} = 0
\label{hatA_5}
\end{equation}
where $m_{0}$ is rest mass of particle, $x_5 = t_r$. 
Let $\hat{F}_{AB} \equiv \partial_{A} \hat{A}_{B} - 
\partial_{B} \hat{A}_{A}$, and A, B runs over 0,1,2,3,5 ( A, B <> 4). 
Energy momentum tensor
\begin{equation}
\hat{T}_{AB} \equiv g_{AB} \hat{F}_{CD} \hat{F}^{CD}/4 - \hat{F}_{A}^{C} \hat{F}_{BC}
\label{1SpinT}
\end{equation}
where A,B run over (0,1,2,3,5, A, B <> 4). Replacing $A_\alpha$ with $\hat{A}_{\alpha}$ in (\ref{6dMetric1S}),
the $\beta4$- components of 6-dimensional Einstein equations become: 
\begin{eqnarray}
\nabla^{\alpha} \, \hat{F}_{\alpha\beta} - m_0^{2} \hat{A}_{\beta} = 0 \; 
\label{6dFieldEq1S}
\end{eqnarray}
And using (\ref{6dFieldEq1S}), the other components of Einstein equations become:
\begin{equation}
G_{AB} = \kappa^2 \hat{T}_{AB} = g_{AB}\kappa^2 \hat{F}_{CD} \hat{F}^{CD}/4 - \kappa^2 \hat{F}_{A}^{C} \hat{F}_{BC}
\label{otherComponent_1Spin}
\end{equation}
which is the definition of $\hat{T}_{AB}$.

Equation(\ref{6dFieldEq1S}) is quantum field equation for massive vector field. 
According to equation(\ref{1SpinT}):
\begin{equation}
 \hat{T}_{44} = \frac{1}{4}\hat{F}_{\alpha\beta} \hat{F}^{\alpha\beta} - \frac{1}{2} m_{0}^{2} \hat{A}_{\alpha} \hat{A}^{\alpha} 
\label{FifthT}
\end{equation}

When the vector field represents 1-spin free boson, equation(\ref{FifthT}) equals zero, i.e 5th dimensional energy momentum tensor
vanished, we get Proca equation:
\begin{equation}
\frac{1}{4}\hat{F}_{\alpha\beta} \hat{F}^{\alpha\beta} - \frac{1}{2} m_{0}^{2} \hat{A}_{\alpha} \hat{A}^{\alpha} = 0 \; \; \; 
\label{6dFieldEquns1S_m}
\end{equation}

\subsection{Gauge transformation invariance for massive gauge field } \label{mass_invarance}

Equation (\ref{6dFieldEquns1S_m}) contains a mass term for gauge field $A_\alpha$. In gauge transformation 
\begin{equation}
A_\mu (x) -> A_\mu '(x) = A_\mu (x) + \partial_\mu  \alpha(x)
\label{gauge_invarance1}
\end{equation}
It is not invariant. But since $m_0$ in (\ref{6dFieldEquns1S_m}) is not a parameter, it comes from $\partial_5 A_\alpha$.
We can rewrite (\ref{6dFieldEquns1S_m}) as 
\begin{equation}
\frac{1}{4}\hat{F}_{\alpha\beta} \hat{F}^{\alpha\beta} + \frac{1}{2} \partial^5 \partial_5 \hat{A}_{\alpha} \hat{A}^{\alpha} = 0 \; \; \; 
\label{6dFieldEquns1S_m_New}
\end{equation}
Since $\partial_5 \alpha(x) = 0$, equation (\ref{6dFieldEquns1S_m_New}) is invariant under gauge transformation (\ref{gauge_invarance1}). Therefore, we 
can derive gauge invariance massive gauge field without Higgs mechanism.

The rest mass of gauge field comes from $e^{\frac{-m_0}{\hbar} t_r}$, this term is based on the fact that the $t_1-t_2$ plane has finite size (metric (\ref{metric_2})).

\subsection{The metric of strong interaction based on $\sigma$ - $\omega$ model  } \label{strong_interaction}

The key condition for Einstein's field equations to work is the assumption that an object is moving in an external field where the size of the source can be neglected. 
This is true for gravitational interactions since the source of gravity, such as a star, is much larger than the object being studied. When examining the interaction 
between two stars, the distance between them is typically large enough that the size effect can be ignored. This condition also applies to electromagnetic interactions in most cases. 
However, for strong interactions, nucleons are very close to each other, and the interaction is very strong. Fortunately, the average potential field approximation for 
strong interactions works well when dealing with many-body problems inside the nucleus. The average field must possess the basic properties of the strong interaction: 
it is repulsive at short distances and attractive at longer distances. 
$\sigma$ - $\omega$ model a.k.a Walecka model\cite{Walecka}\cite{Yao} provides two type of mesons: a scalar meson field $\sigma$ which provides the attractive 
potential, a vector meson field $\omega$ which provides the repulsive potential. The lagrangian of $\sigma$ - $\omega$ model\cite{Walecka}\cite{Yao} is 
\begin{align}
\mathcal{L} &= \overline{\psi} (\gamma_\mu (i\partial^mu - g_v V^\mu) - (M - g_s \phi)) \psi + \frac{1}{2} 
	\partial_\mu \phi \partial^\mu \phi - m_{s}^2\phi^2  \nonumber \\
			& - \frac{1}{4}F_{\mu\nu}F^{\mu\nu} + \frac{1}{2}m_{}^2 V_{\mu} V^{\mu} + \Delta\mathcal{L}
\label{Walecka}
\end{align} 
where $F^{\mu\nu} = \partial^\mu V^\nu - \partial^\nu V^\mu$, $\psi$ is nuclei field, $\phi (V^\mu)$ are scalar (vector) meson fields, M is the mass of nucleon,
$m_s(m_v)$ is the mass of scalar (vector) meson, $g_s (g_v)$ is coupling constant of scalar (vector) meson field and nucleon field.  $\Delta\mathcal{L}$
is renormalization cancellation term. 

The metric of $\sigma$ - $\omega$ model can be written as:
\begin{align}
ds^2 & = g_{\alpha\beta}dx^\alpha dx^\beta - dx^5 dx^5 - (\phi dx^4 + g_v \phi \hat{K}_\alpha dx^\alpha \nonumber \\
     & - \phi (M - g_s \hat{K}_5)dx^5)^2  
\label{interval_charged_halfspin2}
\end{align}
Let  
\begin{align}
& dx^{4'} = dx^4 + g_v \hat{K}_\alpha dx^{\alpha} - (M - g_s \hat{K}_5) dx^5,  \nonumber \\ 
& dx^{\alpha '} = dx^{\alpha}, \; \; \; \;  \; \; dx^{5'} = dx^{5} 
\label{localInertial3}
\end{align}
and 
\begin{align}
& \partial_{\alpha} ' = \partial_{\alpha} - g_v \hat{K}_{\alpha} \partial_{4}  \nonumber \\
& \partial_{4}' = \partial_{4}, \; \; \; \;  \partial_{5}' = \partial_{5} + (M -g_s \hat{K}_5) \partial_4
\label{string_interaction_metric}
\end{align}
where $\partial_5\hat{K}_\alpha = \frac{-m_v}{\hbar} \hat{K}_\alpha$, $\partial_5\hat{K}_5 = \frac{-m_s}{\hbar} \hat{K}_5$  Then equation (\ref{newDirac}) becomes
\begin{equation}
P'\hat{\psi} ' = -i\hbar M ' \partial_4 \psi_s ' 
\label{newDirac_strong}
\end{equation}
where $M ' = -i(M - g_s \hat{K}_5)$, $\psi_s ' = i(M + g_s) \hat{K}_5 \psi$. Expand (\ref{newDirac_strong}), we have 
\begin{equation}
\hbar(i\sigma_\mu \partial^{\mu} - g_v \hat{K}^{\mu})  \hat{\psi} ' - ( m_0 - g_s \hat{K}_5) \psi_s = 0 
\label{newDirac_strong2}
\end{equation}
where $\mu$ runs over 0,1,2,3.
If we let $\hat{K}^{\mu} = V^{\mu}$, $\hat{K}^5 = \phi$, where $V^{\mu}$ ($\phi$) are $\omega$($\sigma$) meson field. Then equation (\ref{newDirac_strong2})
gives the first term of (\ref{Walecka}). 
 
\subsection{Unification of electromagnetic interaction, strong interaction and gravitational interaction} \label{unification1}

We can include electromagnetic, strong interaction and gravitational interaction in the same metric:
\begin{equation}
\begin{split}
ds^2 &= g_{\alpha\beta}dx^\alpha dx^\beta - dx^5 dx^5 - (\phi dx^4 + (g_e \phi A_{\alpha} \\
	 & + g_v \phi \hat{K}_\alpha) dx^\alpha + \phi (M - g_s \hat{K}_5)dx^5)^2  
\end{split}
\label{interval_charged_unified}
\end{equation}
where $\hat{K}^{\alpha} = V^{\alpha}$ is a massive vector meson $\omega$ with mass $m_v$, $\hat{K}^{5} = \phi_s$ is a massive scalar meson $\sigma$ with 
mass $m_s$. $A_\alpha$ is electromagnetic vector. 
Following the transformation process as before, we let
\begin{align}
& dx^{4'} = dx^4 + (g_e A_{\alpha} + g_v \hat{K}_\alpha) dx^{\alpha} - (M_s - g_s \hat{K}_5) dx^5 \nonumber \\
& dx^{\alpha '} = dx^{\alpha}, \; \;  \; \; dx^{5'} = dx^{5} 
\label{localInertial_unified}		
\end{align}
and
\begin{align}
& \partial_{\alpha} ' = \partial_{\alpha} - (g_e A_{\alpha} + g_v \hat{K}_{\alpha}) \partial_{4} \nonumber \\
& \partial_{4}' = \partial_{4},  \; \; \; \; \partial_{5}' = \partial_{5} + (M -g_s \hat{K}_5) \partial_4
\label{string_interaction_metric_unified}
\end{align}
where $\partial_5\hat{K}_\alpha = \frac{-m_v}{\hbar} \hat{K}_\alpha$, $\partial_5\hat{K}_5 = \frac{-m_s}{\hbar} \hat{K}_5$.

The metric (\ref{interval_charged_unified}) contains electromagnetic vector $A_\alpha$ for electromagnetic interaction, massive vector meson 
field $\hat{K}_\alpha$ provides repulsive force of strong interaction, massive scalar meson field $\hat{K}_5$ provides attractive force of 
strong interaction, $\phi$ is the field for nucleon. The nucleon gains its mass from its 5th component M. $\hat{K}_A$ obtains mass from the derivative 
of the 6th dimension. The nucleon derives its charge and coupling constant with mesons from the derivative of the 5th dimension. 
Equation (\ref{string_interaction_metric_unified}) shows that all gauge fields (A and $\hat{K}$) have minimal coupling with the nucleon. 
Once we substitute (\ref{interval_charged_unified}) into (\ref{5dEFE1}), and then choose $g_{\alpha\beta}$  as the metric of gravitational field, we get 
a unified 6-dimensional Einstein equations including gravitational interaction, electromagnetic interaction and strong interaction. 

There is one issue. In gravitational interaction, $\kappa = 4\pi G $ in (\ref{5dEFE1}). But our previous equations, we have : 
\begin{equation}
 -\frac{1}{\psi} \partial_{\alpha}\partial_{\beta}\psi = \kappa T_{\alpha\beta} 
\label{energyMomenum3}
\end{equation}
for plane wave, we get $\frac{1}{\hbar}^2 p_\alpha p_\beta = \kappa T_{\alpha\beta} $.
Thus, it is easy to see  $\kappa = (\frac{1}{\hbar})^2$. We cannot have two different $\kappa$ value if we want to write gravitational interaction 
and quantum level interactions in the same equations. One of the solutions is to introduce scaling factor:
\begin{align}
 \hat{g}_{AB} & -> \hat{g}_{AB} = \Omega^2 \hat{g}_{AB}  \nonumber \\
 \hat{g}^{AB} & -> \hat{g}^{AB} = \Omega^{-2} \hat{g}^{AB} 
\label{scaling1}
\end{align} 
Then
\begin{equation}
 \partial_A ' = \frac{1}{\Omega} \partial_A, \; \; \; \; \; \;  {\partial^A} ' = \Omega \partial^A
\label{scaling2}
\end{equation} 
Anything contains $\partial^{A} \partial_{A}$ won't change, such as Klein-Gordon equations, $p^\alpha p_\beta$, $F_{\alpha\beta} F^{\alpha\beta}$ won't change. 
But the formula of energy-momentum tensor will become: 
\begin{equation}
 -\frac{1}{\psi} \partial_{\alpha} '\partial_{\beta} ' \psi = -\frac{1}{\Omega^2 \psi} \partial_{\alpha} \partial_{\beta} \psi \kappa T_{\alpha\beta} 
\label{energyMomenum4}
\end{equation} 
Let $\Omega =\frac{\sqrt{4\pi G}}{\hbar}$, we get $\kappa = 4\pi G $. It is interesting to see that the scaling factor is the division between
 macroscopic scale constant G and the microscopic scale constant $\hbar$.

\subsection{Wave-packet collapse and Causality} \label{WavePacketCollapse}

\begin{figure}
  \includegraphics[scale=1.0]{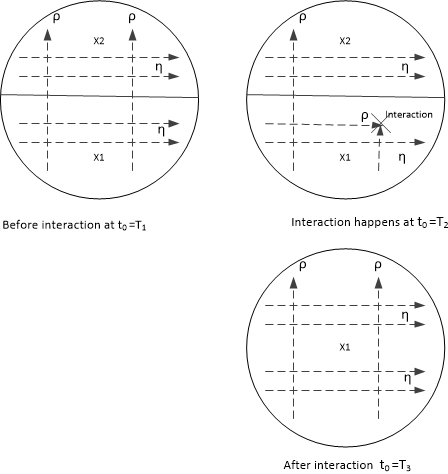}
  \caption{Time branch changes before and after interaction}
  \label{fig:collapse}
\end{figure}
 
Wave-packet collapse is a crucial factor in measurement theory in quantum physics. But so far we don't have a good explanation of wave-packet collapse.
Mathematically it is an operator projecting N different states to a single state: $(|S_j >)< \Sigma_{i} S_i|$. This operator is non-Hermitian. The 
Hermitian requirement in quantum physics is based on that the probability of finding a value of  an operator O is $\psi^*O\psi$. In 3-d time, we know the probability
is proportional to $|\Delta t_1 \times \Delta t_2|$,  $\Delta t_1$ is not necessarily equals $\Delta t_2$. in other words, if $\eta$ motion 
proportional to $\psi$, $\rho$ motion is not necessarily proportional to $\psi^*$. Fig \ref{fig:collapse} shows the change of time branches before and after 
interaction. In Fig \ref{fig:collapse}, the $t_1-t_2$ plane is divided into two branches. 
At different time branches, the particle has different space region X1 (X2).  Timeline $\eta$ parallel to the branch separation line, 
timeline $\rho$ perpendicular to the separation line. Therefore, it goes through the line.
$\eta$ and $\rho$ are not symmetrical. When another identical particle boson exists, due to the symmetrical between $t_1$ and $t_2$, 
it is likely the $\rho_2$ (second particle) is parallel to the separation line and $\eta_2$ is perpendicular to the line. So when we look at two particle
cases, timeline $\eta$ and $\rho$ are symmetrical. When there are many particles involved, the difference between $\eta$ and $\rho$ will be averaged out. 
Then probability = $\psi^*O\psi$ is a valid approximation. But wave-packet collapse only happens on each single particle, in such event,
the difference between $\eta$ and $\rho$ could be important. In Fig \ref{fig:collapse}, the particle's timeline $\rho$ loops over X1 and X2 branch. The separation of 
the branches could be caused by some slow processes acting on the particle's paths over time as we discussed in section \ref{TIMEBRANCHS}, which caused 
discontinuity at the separation line on $\rho$ motion. When a measurement happens locally 
only at X1 branch (the bottom half of the Fig.), it creates a new discontinuity which could wipe out the discontinuities created historically . The easiest way to 
describe the process is to consider the particle as being annihilated and then recreated at the X1 branch. This recreation would wipe out the historical information. 
When the new particle's timeline moves to the top region, its physical properties will be based on the properties at the time the new particle was born. 
As a result, the top half of the time area is simply an extension of the physical properties of X1 from the bottom half.

Timeline on $t_1-t_2$ plane forms time loop. We could not have causality on a time loop since we cannot distinguish which event happened first in a 
loop. Furthermore, the particle can stay in different locations at the same clock time, which is against the causality in relativity. Time loop only happens 
on $t_1-t_2$, there is no time loop on $t_0$, no time loop on motion $\tau$. The causality is related to the order of events. An event is related to 
an observable measurement, which needs to interact with the particle. If the interaction acts on the particle's timelines on $t_1-t_2$ as the whole, 
it won't detect time loop or non-locality because we couldn't tell which location the interaction happened at. If the interaction acts on a small area 
of the particle's timelines on $t_1-t_2$, it will cause the 
wave-packet collapse, which destroys time loop and non-locality on $t_1-t_2$ plane. Due to the wave-packet collapse, we cannot measure the two events 
happening on a time loop on $t_1-t_2$ timelines, nor can we find a particle showing up at two 
space locations at the same time.  Therefore, causality still holds in 3-d time physics. In macroscopic world, an object is composed by large number of particles,
the wavepacket-collapse is caused by the interactions among those particles happening at any space-time point. Therefore, the causality holds and we see only $\tau$ 
motion in macroscopic world.
   
\section{Discussion and Conclusion} \label{discussion}

In this paper, a framework of 3-d time physics is created. We use three separate motions to reproduce wave-function of 0-spin, 1-spin and 1/2-spin particle. 
Each motion associates with a timeline vector in 3-d time. For 0-spin and integer spin particle, the phase of wave-function is direction of 
time vector in $t_1-t_2$ plane. For 1/2-spin particle, time vector has three dimensional directions in time coordinates. The two-components spinor 
of 1/2-spin particle is built based on three directions of time vector. For a free particle, motions $\eta$, $\rho$ have constant velocities. 
Motion $\tau$ provides an oscillation on all timelines due to energy of the particle. The field equations are based on 6-dimensional KK equations. 
When dealing with interactions, we write a 3+3 space-time metric such that the interactions can be included in partial derivative through coordinate
transformation. The new partial derivative is the minimal coupling.  The mass of spinless and integer spin particle comes from boundary condition of the $t_1-t_2$ plane. 
The mass of 1/2-spin particle is coming from a scalar field in space-time metric on 6th dimension. 
The charge, mass and coupling constant with scalar (vector) meson come from the derivative of a time dimension. Gauge field obtains the mass without breaking gauge invariance. 

There are two major issues on 5-dimensional KK theory. First, it can only deal with massless fields. Second, it cannot provide the convincing proof
of the existence of 5th dimension. The extra dimension in our 3+3 KK equations covered the first issue. For the second issue, quantum physics itself is the evidence 
to show the existence of 2nd and 3rd  time dimensions. The point to use 6-dimensional Einstein equation to unify interaction is to show that all the interactions 
stem from the same root: the distorted space-time geometry. We are not only able to include gravitational interactions, electromagnetic interactions and massive mesons in 
the same 6-dimensional Einstein equations, but also show that Einstein equations can explain the motion of the free quantum particles. 

The 3-d time shows cylindrical symmetry. $t_0$, $t_1$, $t_2$ are not equal. 
One possible explanation is that before the Big Bang, the three dimensions of time were nearly equal, forming a confined 'time ball.' However, for some reason, 
one of these time dimensions broke the symmetry and became the primary time of our universe. This symmetry breaking could have caused the Big Bang, 
or it might have been a result of it. 
   
The main reason for 3-d time physics is to have a clear picture of how a single particle moves in quantum physics. One of the differences between 3-d time 
motion and quantum physics is that $\eta$ and $\rho$ are two separate motions. Therefore, we should study $\eta \times \rho$  instead of $\psi^* \psi$. 
Another new aspect is that the velocity should be a $3 \times 3$  matrix. With certain conditions, this $3 \times 3$ matrix could be related to SU(3).
Another implementation could be to deal with the infinity caused by the divergence of the path integral 
on loop diagram. There are two methods of renormalization. 1) Introduce a shape factor to cut off the range of the integration. 2) Reduce the dimension of 
integration from d dimensions to d-2. Since $t_1-t_2$ plane is finite, it is possible to implement the finite metric of the $t_1-t_2$ to cut off the 
range of the integration. 

Starting from 1999, I authored a few papers regarding interpreting quantum physics using 3-d time\cite{xchen}, and 3-d time on unification theory\cite{xchen2}.
Those papers didn't give a clear picture of how the particle moves in 3-d time. This paper is trying to layout the fundamental concepts of 3-d time physics.

\section{Acknowlegement} \label{Ack}

I'd like to dedicate this paper to professor Y. Yao and S.S. Wu. When I was just a junor college student, I proposed the 3-dimensional time idea 
to Professor Yao. Her encouragment at that time helped me work through the difficulties of theory in the past many years .


\begin{thebibliography}{99}

\bibitem{Kal21} T. Kaluza, {\it Zum Unit\"atsproblem der Physik},
                 Sitz. Preuss. Akad. Wiss. Phys. Math. K1 (1921) 966. 

\bibitem{Overduin}  J. M. Overduin, P.S.Wilson, {\it Kaluza-Klein Gravity},
		http://arxiv.org/abs/gr-qc/9805018

\bibitem{Kle26a} O. Klein, {\it Quantentheorie und f\"unfdimensionale 
                  Relativit\"atstheorie },
                  Zeits. Phys. 37 (1926) 895.                  		
				  
\bibitem{FUETER}	R. FUETER,  Comment. Math. Helv. 7 (1935), 307-330.					  

\bibitem{Sudbery} A. Sudbery {\it Qaternionic Analysis}, https://dougsweetser.github.io/Q/Stuff/pdfs/Quaternionic-analysis.pdf				  
	

\bibitem{Edmonds} J. Edmonds Jr.,{\it Maxwell’s eight equations as one quaternion equation},
		Am. J. Phys. 46, 430–431 (1978) https://doi.org/10.1119/1.11316

\bibitem{Walecka} J.D. Walecka,{\it A Theory of highly condensed matter},
		Ann. Phys. 83 (1974) 491 https://doi.org/10.1016/0003-4916(74)90208-5
		
\bibitem{Yao} Y. Yao, X. Chen, S. S. Wu,  {\it Contribution of negative energy sea in the $\sigma - \omega$ model},
        	High Energy Phys. Nucl. Phys.(China), vol.22(1998) No.3. 247

\bibitem{xchen} X. Chen , {\it A new interpretation of quantum theory - Time as hidden variable},
		https://doi.org/10.48550/arXiv.quant-ph/9902037
		
\bibitem{xchen2} X. Chen , {\it Modified Kaluza-Klein Theory, Quantum Hidden Variables and 3-Dimensional Time},
		https://arxiv.org/pdf/quant-ph/0501034

\end{thebibliography}
\end{document}